\documentclass[aps,prd,12pt,onecolumn,preprintnumbers,floatfix,amsmath,amssymb,nofootinbib]{revtex4-2}

\usepackage{amsfonts,bm}
\usepackage{booktabs,multirow,array,float,graphicx,enumitem}
\usepackage[
    colorlinks=true,
    linkcolor=blue,
    citecolor=blue,
    urlcolor=blue,
    bookmarks=true,
    bookmarksnumbered=true,
    unicode=true,
    breaklinks=true,
    pdfstartview=FitH,
    hyperindex=true
]{hyperref}
\makeatletter
\providecommand\@cite[1]{[#1]}
\providecommand\@biblabel[1]{#1.}
\makeatother

\begin{document}

\title{Photon Spheres and Shadows of Covariant Loop Quantum Black Holes in the general $\mu$-scheme}
\author{Yu Han}
\email{hanyu@xynu.edu.cn}
\affiliation{College of Physics and Electrical Engineering, Xinyang Normal University, 464000 Xinyang, China}
\author{Meng Liu}
\affiliation{College of Physics and Electrical Engineering, Xinyang Normal University, 464000 Xinyang, China}
\author{Yong-Zhuang Li}
\affiliation{School of Science, Jiangsu University of Science and Technology, 212100 Zhenjiang, China}

\begin{abstract}
We investigate black hole shadows and photon sphere properties for two families of covariant quantum-corrected black-hole metrics (hereafter called BH-I and BH-II) formulated within a general $\mu$-scheme, parameterised by a power-law exponent $s$ and an amplitude $\xi$. The extension to general (including non-integer) $s$ is a phenomenological interpolation between the $\mu_0$-scheme ($s=0$) and $\bar\mu$-scheme ($s=1$) and reveals a rich phenomenology masked when $s$ is usually fixed to 1 in previous literature. For BH-I, the photon sphere exists for all parameter values and exhibits an exact cancellation at $s=1$ where the coordinate location $r_{\rm ph}=3M$ is restored for any $\xi$. For BH-II, both the horizon and the photon sphere exhibit critical curves in the $(\xi,s)$ parameter plane; the photon sphere disappears when $\xi$ exceeds a closed-form critical value $\xi_c^{\rm PS}(s)$. We prove analytically that the photon sphere is always unstable ($\lambda_{\rm ph}>0$) throughout the physical parameter space of both metrics, and derive closed-form expressions for the Lyapunov exponent. Systematic parameter scans reveal that for fixed $\xi$ the shadow radius decreases monotonically with $s$ for BH-II and non-monotonically for BH-I. We derive small-parameter analytic expansions for the photon-sphere radius, shadow radius, and Lyapunov exponent, and demonstrate that a single shadow measurement suffers an observational degeneracy in the two-dimensional $(\xi,s)$ space; the degeneracy can be broken by a simultaneous measurement of the Lyapunov exponent. Using Event Horizon Telescope (EHT) measurements, we derive constraints on the $(\xi,s)$ parameter space and find that BH-II is constrained roughly $2$--$4$ times more tightly than BH-I.
\end{abstract}

\maketitle

\section{Introduction}\label{sec:introduction}

Although the predictions of gravitational waves and black holes by general relativity (GR) have been firmly supported by observational evidence \cite{Abbott2016, EHT2019M87, EHT2022SgrA}, according to the singularity theorems \cite{Penrose1965}, the gravitational collapse of massive stars in GR inevitably leads to spacetime singularities, this suggests that GR may break down in extreme regimes. To address these issues, various approaches to quantum gravity have been proposed. Among them, loop quantum gravity (LQG), as a non-perturbative and background-independent approach to quantum gravity, has attracted considerable attention and led to fruitful developments \cite{Ashtekar2004, Rovelli2004, Ashtekar2006a}. LQG directly quantizes spacetime, positing that spacetime is composed of discrete units, thereby avoiding the problem of infinite curvature at singularities. In the last fifteen years, the quantum extension of classical Kruskal spacetime has been explored within this framework in various literature (see, for instance, \cite{Gambini2013, Ashtekar2018, Bodendorfer2019, Bodendorfer2021, Bojowald2020, Husain2022}).

Although LQG has made remarkable progress in resolving classical singularities, whether its canonical formulation preserves covariance after the introduction of quantum effective corrections has remained an important issue, prompting extensive efforts to address this issue \cite{Bojowald2015Covariance, Brahma2018, Gambini2022, Bojowald2022, AlonsoBardaji2022, AlonsoBardaji2022b, AlonsoBardaji2024, Han2024}. Recently, within the Hamiltonian framework of effective quantum gravity inspired by LQG, the conditions required for covariance have been rigorously derived in the static and spherically symmetric case, and several covariant quantum-corrected black hole solutions have been constructed \cite{Zhang2025LQLett, Zhang2025NoCauchy, Belfaqih2025, Zhang2025CovDyn, Yang2025Electrovacuum}, providing a new platform for studying quantum effects while preserving covariance.

In the LQG framework, quantum-corrected black hole metrics arise from the effective polymerization of the Ashtekar connection variables or the extrinsic curvature variables in the Kantowski-Sachs minisuperspace \cite{Munch2023, Zhang2023LQBH}. There are different ways to polymerize the Ashtekar connection or the extrinsic curvature variable, some polymerization scheme does not involve the black hole radius while some does, very similar to the polymerization schemes called the $\mu_{0}$-scheme or $\bar{\mu}$-scheme used in loop quantum cosmology and LQG \cite{Ashtekar2006Improved, Bojowald2008LRR, HanLiu2020Improved}; in the following text, we also adopt these two terms and generalize the $\mu_{0}$-scheme or $\bar{\mu}$-scheme in loop quantum black hole. The general $\mu$-scheme is inspired by the lattice refinement program in loop quantum cosmology \cite{Bojowald2008LRR, Bojowald2007Lattice, Bojowald2006Inhom}. In the full theory of loop quantum gravity, the Hamiltonian constraint operator creates new vertices on the spin-network state as the universe expands, leading to a dynamical refinement of the underlying lattice. In the cosmological context this refinement is parameterized by the polymerization scale $\mu\propto V^{m}$, where $V$ is the spatial volume and $m$ is the refinement exponent. The standard $\mu_{0}$-scheme corresponds to fixed lattice spacing, while the $\bar{\mu}$-scheme corresponds to a lattice number proportional to volume. In the black-hole context the radius $r$ plays the role of the effective volume scale, and the general $\mu$-scheme with $\mu\propto r^{-s}$ maps onto a similar structure: the exponent $s$ plays a role analogous to the lattice-refinement exponent, controlling how rapidly the polymerization scale adapts to the local geometry. From this perspective the cases $s=0$ and $s=1$ correspond, respectively, to the fixed-lattice and volume-adaptive refinement patterns familiar from loop quantum cosmology. The generalization to non-integer $s$ can be viewed as a phenomenological extension that interpolates between these two well-motivated limits, much as general refinement models with different lattice-refinement exponents $m$ have been studied in the cosmological setting to explore the robustness of physical predictions. However, most existing studies have focused on the special case $s=1$ (the $\bar{\mu}$-scheme), leaving the full $(\xi,s)$ parameter space largely unexplored.

Black hole shadows encode spacetime geometry near the photon sphere and provide a powerful observational probe of strong-field regime gravitational effects \cite{Falcke2000, Johannsen2012}. The Event Horizon Telescope (EHT) has now resolved the shadow angular diameter for both M87* \cite{EHT2019M87} and Sgr~A* \cite{EHT2022SgrA}, opening a new window for testing strong-field gravity. With further advances in observational techniques, quantum gravitational effects in the strong-field regime are expected to become detectable with next-generation telescopes.

In this paper we study photon sphere and shadow properties of the BH-I and BH-II metrics within the general $\mu$-scheme. The paper is organized as follows. In Sec.~\ref{sec:metric} we introduce the two metric families (BH-I and BH-II), establish the geometric framework, and analyze their horizon structure. In Sec.~\ref{sec:shadows} we analyze the photon sphere existence and uniqueness for both BH families (Sec.~\ref{subsec:photon-shadow-BH-I} and Sec.~\ref{subsec:photon-shadow-BH-II}), derive small-parameter analytic expansions that give insight into the $(\xi,s)$ dependence (Sec.~\ref{subsec:small-param}), prove that the photon spheres are unstable and derive closed-form expressions for the Lyapunov exponent, obtain observational constraints from EHT shadow-diameter measurements and discuss the observational degeneracy inherent in shadow-only measurements (Sec.~\ref{subsec:eht-constraints}). In Sec.~\ref{sec:discussion} we summarize our findings and discuss future prospects. Throughout this paper, we adopt geometric units $G = c = 1$.

\section{Metric Background}\label{sec:metric}

We present the quantum-corrected metric and establish the geometric framework for subsequent analyses. The metric is a static, spherically symmetric spacetime of the form
\begin{equation}\label{eq:general_metric}
d s^2 = -f(r)\,dt^2 + \frac{1}{g(r)}\,dr^2 + r^2\left(d\theta^2 + \sin^2\theta\,d\phi^2\right),
\end{equation}
In loop quantum gravity, the holonomy of extrinsic curvature $k$ around a small square with edge length $\zeta$ is given by
\begin{equation}\label{holonomy}
k\rightarrow \frac{\sin(\mu k)}{\mu},\qquad~ \mu=\frac{\zeta}{l}
\end{equation}
in which $l$ is an auxiliary parameter with the dimension of length and introduced here to make $\mu$ dimensionless, because $k$ is dimensionless and $\mu\!k$ should also be dimensionless. In the black-hole context a convenient choice is $l=aGM$ with $a$ an unspecified dimensionless coefficient (other choices are also allowed \cite{Ashtekar2018}).

Generally speaking, $\zeta$ may be a constant or vary with $r$. In this article, we call the case $\mu= \zeta_0/(aGM)$ the $\mu_0$-scheme in loop quantum black hole, which is similar to the $\mu_0$-scheme in loop quantum cosmology. We can also choose $\zeta(r)=\zeta_0\bigl(bGM/r\bigr)$ in which $b$ is also an unspecified dimensionless parameter such that the terms inside the bracket is dimensionless and $\zeta(r)$ has the same dimension as $\zeta_0$, under this choice, we have $\mu=\zeta(r)/(aGM)=(b/a)\cdot(\zeta_0/r)$, and we call this choice the $\bar{\mu}$-scheme in loop quantum black hole, which mimics the $\bar{\mu}$-scheme in loop quantum cosmology. The $\bar{\mu}$-scheme has been widely used in the current research of loop quantum black holes \cite{Zhang2020, AlonsoBardaji2022b, AlonsoBardaji2024, AlonsoBardaji2022, Belfaqih2025, Zhang2025NoCauchy}. In this article, we consider a more general parameterization
\begin{equation}
\zeta(r)=\zeta_0 \left(\dfrac{b\,GM}{r}\right)^{s},
\end{equation}
where $s$ is an unspecified positive constant; note that $s=0$ corresponds to the $\mu_0$-scheme and $s=1$ corresponds to the $\bar{\mu}$-scheme. Hence, in the general $\mu$-scheme, we have
\begin{equation}\label{param}
\mu=\left(\dfrac{\zeta_0}{a\,GM}\right)\left(\dfrac{b\,GM}{r}\right)^{s},
\end{equation}
defining $\xi\equiv\zeta_0\,b^s/a$, the above expression can be rewritten as
\begin{equation}\label{param2}
\mu=\left(\dfrac{\xi}{GM}\right)\left(\dfrac{GM}{r}\right)^{s},
\end{equation}
note that $\xi$ inherits the dimension of length from $\zeta_0$ and becomes an undetermined constant since the unspecified constants $\zeta_0$, $a$ and $b$ have been absorbed into it.

\subsection{Metric Functions}\label{subsec:line_element}

In this work we consider two families of quantum-corrected black holes, both derived from the holonomy correction~\eqref{param} but differing in the way the quantum term is incorporated into the metric. We refer to them as BH-I and BH-II.

The first metric preserves the Schwarzschild symmetry $f(r) = g(r)$:
\begin{equation}\label{eq:BH1_f}
f_{\rm I}(r) = g_{\rm I}(r) = 1 - \frac{2M}{r} + \left(\frac{\xi}{M}\right)^{2}\!\left(\frac{M}{r}\right)^{2s}\!\left(1 - \frac{2M}{r}\right)^2,
\end{equation}
where $\xi$ is a quantum correction parameter with the dimension of mass (so $\xi/M$ is dimensionless), and the $(1-2M/r)^2$ factor ensures the correction vanishes at the Schwarzschild horizon $r = 2M$. This metric is derived using the effective mass $M_{\rm eff}^{(1)}$ with a certain polymerization of the extrinsic curvature proposed in Ref.~\cite{Zhang2025LQLett} under the covariance formalism. In this work we restrict to $s > 1/2$, so that the quantum correction decays faster than the Newtonian $M/r$ term at large radii; explicitly, $f_{\rm I}(r)=1-2M/r+O((M/r)^{2s})$ as $r\to\infty$, which guarantees the ADM mass is well defined and equals $M$. At the marginal value $s=1/2$, the correction would decay as $(M/r)^{1}$, interfering with the definition of the ADM mass; we therefore exclude this point.

A second family arises when the quantum correction enters as a pure power law without the $(1-2M/r)^2$ prefactor:
\begin{equation}\label{eq:BH2_f}
f_{\rm II}(r) = g_{\rm II}(r) = 1 - \frac{2M}{r} + \left(\frac{\xi}{M}\right)^{2}\!\left(\frac{M}{r}\right)^{2s}\!\left(\frac{2M}{r}\right)^{2},
\end{equation}
where $\xi$ is a quantum correction parameter with the dimension of mass. This metric can be derived using an alternative effective mass function with a different polymerization of the extrinsic curvature from that in Ref.~\cite{Zhang2025LQLett}, and in the Appendix we will give a detailed derivation of this metric using the new effective mass function. For BH-II we allow $s \geq 0$; the quantum correction decays as $r^{-(2s+2)}$, which is always faster than the Newtonian $M/r$ term for any $s \geq 0$, so the ADM mass is well defined and equals $M$ for all parameter values. Notably, for $s=0$ the metric reduces to $f_{\rm II}=1-2M/r+4\xi^{2}/r^{2}$, which is formally identical to the Reissner--Nordstr\"om metric with effective charge $Q=2\xi$ (i.e. $Q^{2}=4\xi^{2}$). For $s=1$, this metric reduces to the Quantum Oppenheimer--Snyder model proposed using loop quantum cosmology in Ref.~\cite{Zhang2022LQC}; this provides a familiar limiting case and a useful benchmark for testing our results. The orbital dynamics of BH-II are studied in detail in Sec.~\ref{sec:shadows}.

\subsection{Horizon Structure of BH-I}\label{subsec:horizon-bh1}

For BH-I the metric factorises as $f_{\rm I}(r)=(1-2M/r)\,Q(r)$ with
\begin{equation}
Q(r)\equiv 1+\left(\frac{\xi}{M}\right)^{2}\!\left(\frac{M}{r}\right)^{2s}\!\left(1-\frac{2M}{r}\right).
\end{equation}
Since $Q(2M)=1>0$ and $Q(r)\to-\infty$ as $r\to0^{+}$ (the negative term $-2M/r$ dominates at small $r$), the intermediate value theorem guarantees at least one root of $Q(r)$ in $(0,2M)$. To see that this root is unique, write $x\equiv r/M$ and $Q(x)=1+a\,x^{-2s}(1-2/x)$ with $a=(\xi/M)^{2}>0$; then $Q'(x)=a\,x^{-2s-2}[-2sx+(4s+2)]$. The bracket is positive for all $x<2+1/s$, and since $2+1/s>2$ for every $s>0$, we have $Q'(x)>0$ throughout $(0,2)$. Thus $Q$ is strictly increasing on $(0,2M)$, so it crosses zero exactly once; that unique root is the inner Cauchy horizon $r_{\rm in}\in(0,2M)$. The outer root comes from the first factor and is fixed at $r_{\rm out}=2M$ for all $(\xi,s)$. The surface gravity $\kappa=\frac12 f'(r_{\rm out})=1/(4M)$ is independent of both $\xi$ and $s$, identical to Schwarzschild. No extremal solution exists because $f_{\rm I}'(2M)=1/(2M)>0$ everywhere. As $\xi\to0$ the inner horizon shrinks to $r_{\rm in}\to0$; at $s=1$ and $\xi=M$ one finds $r_{\rm in}=M$.

Since the inner-horizon satisfies the equation $x_{\rm in}^{2s+1}+a\,x_{\rm in}-2a=0$, treating $x_{\rm in}$ as a function of $s$ gives $\mathrm{d}x_{\rm in}/\mathrm{d}s=-2x_{\rm in}^{2s+1}\ln x_{\rm in}/[(2s+1)x_{\rm in}^{2s}+a]$, so $x_{\rm in}(s)$ decreases for $x_{\rm in}>1$ ($\xi/M>1$) and increases for $x_{\rm in}<1$ ($\xi/M<1$). In all cases $x_{\rm in}\leq 1$ (i.e. $r_{\rm in}\leq M$) iff $\xi/M\leq 1$; for $\xi/M>1$ the inner horizon can approach the outer horizon as $\xi\to\infty$.

For fixed $s$, treating $x_{\rm in}$ as a function of $a$ gives $\mathrm{d}x_{\rm in}/\mathrm{d}a=(2-x_{\rm in})/[(2s+1)x_{\rm in}^{2s}+a]>0$, so $x_{\rm in}$ increases monotonically with $\xi/M$ from $0$ as $\xi\to0$ toward $2$ (approaching the outer horizon) as $\xi\to\infty$. Unlike BH-II, no critical curve exists---the inner horizon is always present.

\subsection{Horizon Structure of BH-II}\label{subsec:horizon-bh2}

For BH-II the horizon equation $f_{\rm II}(r_{\rm h})=0$ is equivalent to $P(r_{\rm h})=0$ with
\begin{equation}
P(r)\equiv r^{2s+2}-2M\,r^{2s+1}+4\left(\frac{\xi}{M}\right)^{2}M^{2s+2}.
\end{equation}
Since $P(0)=C>0$ (with $C$ the constant term) and $P(r)\to+\infty$ as $r\to\infty$, the number of positive roots is controlled by the minimum of $P(r)$. The derivative $P'(r)=r^{2s}[(2s+2)r-2M(2s+1)]$ vanishes at the critical radius
\begin{equation}\label{eq:rh-c}
r_{\rm h}^{c}=\frac{2s+1}{s+1}\,M,
\end{equation}
which satisfies $M\leq r_{\rm h}^{c}<2M$ for all $s\geq0$. At this point
\begin{equation}
P(r_{\rm h}^{c})=C-\frac{M\,(r_{\rm h}^{c})^{2s+1}}{s+1},
\end{equation}
so two distinct roots exist if and only if $P(r_{\rm h}^{c})<0$, i.e.\ $\xi<\xi_c^{\rm h}(s)$ with the critical curve
\begin{equation}\label{eq:xi_c_h}
\frac{\xi_c^{\rm h}(s)}{M}=\frac{(2s+1)^{s+\frac12}}{2\,(s+1)^{s+1}}
\qquad(s\geq0).
\end{equation}
In particular, $\xi_c^{\rm h}/M=1/2$ for $s=0$ (the Reissner--Nordstr\"om-like limit) and $3\sqrt{3}/8\approx0.650$ for $s=1$ (the $\bar{\mu}$-scheme).

\begin{figure}[htbp]
\centering
\includegraphics[width=0.6\textwidth]{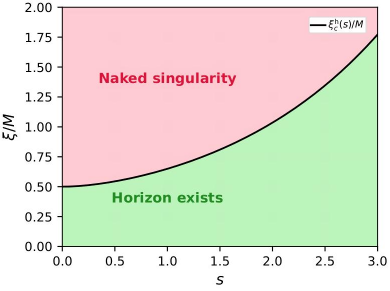}
\caption{Horizon-existence boundary for BH-II in the $(s,\xi/M)$ plane. The solid curve is the analytic critical curve $\xi_{c}^{\rm h}(s)$ from Eq.~\eqref{eq:xi_c_h}. The green (light) region: $\xi<\xi_{c}^{\rm h}(s)$, where a regular event horizon exists. The red (dark) region: $\xi>\xi_{c}^{\rm h}(s)$, corresponding to a naked singularity.}
\label{fig:horizon_critical_curve}
\end{figure}

Figure~\ref{fig:horizon_critical_curve} shows the critical curve $\xi_{c}^{\rm h}(s)$ in the $(s,\xi/M)$ plane, separating the horizon-existence (green) from the naked-singularity (red) regime.

When the horizon condition holds there are two horizons: an inner Cauchy horizon $r_{\rm in}\in(0,r_{\rm h}^{c})$ and an outer event horizon $r_{\rm out}\in(r_{\rm h}^{c},2M)$. The upper bound $r_{\rm out}<2M$ follows from $P(2M)=C>0$ together with the monotonicity of $P$ on $(r_{\rm h}^{c},\infty)$; the quantum correction shifts the event horizon inward relative to Schwarzschild. As $\xi$ increases at fixed $s$, $r_{\rm in}$ rises and $r_{\rm out}$ falls until they meet at the extremal point where $\xi=\xi_c^{\rm h}(s)$ and both horizons coalesce at $r_{*}=r_{\rm h}^{c}$.
For $\xi>\xi_c^{\rm h}(s)$ no horizon exists and the spacetime contains a naked singularity. The use of the term ``singularity'' is justified by the behaviour of curvature invariants: as $r\to0^{+}$ the dominant term of the metric function is $f_{\rm II}\sim 4\xi^{2}M^{2s}r^{-(2s+2)}$, so the Kretschmann scalar diverges as $\mathcal{K}\sim r^{-(4s+8)}$ for any $s\geq0$. The divergence is therefore a genuine curvature singularity of the effective metric (not a coordinate artefact), in contrast to the non-singular bounce cores found in some other effective LQG models; whether the full quantum theory resolves it lies outside the scope of the effective description. Such configurations are excluded by the cosmic censorship hypothesis; in our analysis we restrict to $\xi<\xi_c^{\rm h}(s)$ and treat the critical curve as the boundary of the physically admissible parameter space.

\begin{figure}[t]
\centering
\includegraphics[width=0.9\textwidth]{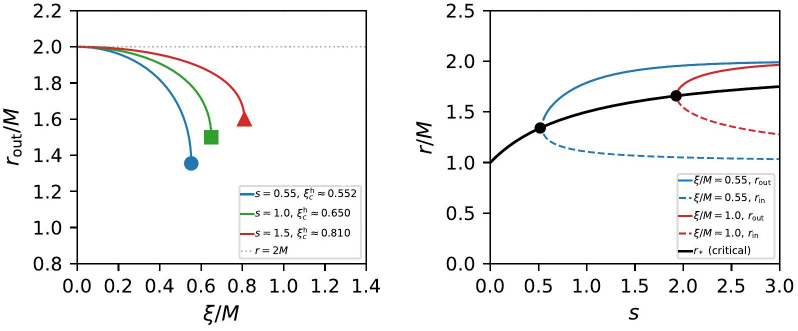}
\caption{BH-II horizon structure. \textbf{(a)} $r_{\rm out}$ vs $\xi/M$ for $s=0.55$ (blue), $s=1.0$ (green) and $s=1.5$ (red). Each curve terminates at its critical value $\xi_c^{\rm h}(s)$ (filled circles and triangle). \textbf{(b)} Outer and inner root locations vs $s$ for $\xi/M=0.55$ (blue) and $\xi/M=1.0$ (red). The solid curves are $r_{\rm out}$ (stable), the dashed curves are $r_{\rm in}$ (unstable), and the thick black curve shows the critical radius $r_*$ where $r_{\rm out}=r_{\rm in}$.}
\label{fig:BH2-rout-and-roots}
\end{figure}

Writing $x\equiv r_{\rm h}/M$ and $a\equiv(\xi/M)^{2}$, the horizon condition becomes $\tilde{P}(x)\equiv x^{2s+2}-2x^{2s+1}+4a=0$. By the implicit function theorem
\begin{equation}
\frac{\partial x}{\partial a}=-\frac{4}{x^{2s}\bigl[(2s+2)x-2(2s+1)\bigr]},
\end{equation}
which is positive for $x_{\rm in}<x_{c}$ and negative for $x_{\rm out}>x_{c}$; hence $r_{\rm in}$ ($r_{\rm out}$) increases (decreases) monotonically with $\xi/M$, from $0$ ($2M$) to the coalescence radius $r_{\rm h}^{c}$. Similarly,
\begin{equation}
\frac{\partial x}{\partial s}=-\frac{2x(x-2)\ln x}{(2s+2)x-2(2s+1)}.
\end{equation}
For the outer horizon ($x_{\rm out}\in(x_{c},2)$) this is positive, so $r_{\rm out}$ increases with $s$ toward $2M$; for the inner horizon $\partial x_{\rm in}/\partial s>0$ when $x_{\rm in}<1$ but changes sign for $x_{\rm in}>1$, and in all cases $r_{\rm in}\leq\min(r_{\rm h}^{c},M)$ for $s\geq0$.

Figure~\ref{fig:BH2-rout-and-roots} shows the horizon locations. The left panel displays $r_{\rm out}$ versus $\xi/M$ for three representative values of $s$: each curve decreases monotonically from the Schwarzschild value $2M$ and terminates at the critical point $(\xi_c^{\rm h}, r_*)$ where the two roots coalesce. The critical value $\xi_c^{\rm h}(s)$ rises with $s$, reflecting that faster-decaying quantum corrections permit larger amplitudes before the boundary disappears. The right panel shows both roots versus $s$ for fixed $\xi/M$.

\section{Photon Spheres and Black Hole Shadows}\label{sec:shadows}

Black hole shadows encode the spacetime geometry near the photon sphere and provide a powerful probe of strong-field gravitational effects \cite{Konoplya2025,Chen2025,Lin2024,Perlick2022}. Previous studies of LQG-inspired black-hole shadows have largely targeted specific models or the $\bar{\mu}$-scheme alone \cite{Konoplya2025,Chen2025,Lin2024}. The present work generalises these analyses in three directions: (i) we cover the full two-dimensional $(\xi,s)$ parameter space of the general $\mu$-scheme, revealing phenomena---the exact cancellation, the critical curves, and the non-monotonic $s$-dependence---invisible at fixed $s$; (ii) we establish the photon-sphere instability analytically over the entire admissible parameter space; and (iii) we derive closed-form critical curves and small-parameter expansions that make the $(\xi,s)$ dependence of all observables explicit, and impose EHT constraints on the parameter space.

The photon sphere radius $r_{\rm ph}$ is the radius of the circular photon orbit around the black hole, and the shadow radius $R_{\rm sh}$ is the apparent radius of the black hole shadow as seen by a distant observer. For a static spherically symmetric metric with $f(r)=g(r)$, the motion of photons in the equatorial plane is governed by the null geodesic equation ($\theta=\pi/2$). The metric possesses two Killing vector fields, the time-like field $(\partial/\partial t)^{a}$ and the axial field $(\partial/\partial\phi)^{a}$, giving rise to two conserved quantities, namely the photon energy $E$ and the angular momentum $L$:
\begin{equation}
E = -g_{ab}\left(\frac{\partial}{\partial t}\right)^{a}\!\dot x^{b} = f(r)\,\dot t,\qquad
L = g_{ab}\left(\frac{\partial}{\partial\phi}\right)^{a}\!\dot x^{b} = r^{2}\,\dot\phi,
\end{equation}
where an overdot denotes differentiation with respect to an affine parameter $\lambda$. The impact parameter is defined as the ratio
\begin{equation}
\label{eq:impact-parameter}
b \equiv \frac{L}{E},
\end{equation}
which uniquely labels each photon trajectory. Dividing the null condition by $L^{2}$ (equivalently, rescaling the affine parameter by $L$), the radial motion takes the effective-potential form
\begin{equation}
\label{eq:null-radial}
\dot r^{2} + V_{\rm eff}(r) = \frac{1}{b^{2}},\qquad V_{\rm eff}(r) \equiv \frac{f(r)}{r^{2}}.
\end{equation}
Circular photon orbits satisfy $V_{\rm eff}'(r_{\rm ph})=0$, which determines $r_{\rm ph}$ independently of $b$. The critical impact parameter follows from $V_{\rm eff}(r_{\rm ph})=1/b_{\rm ph}^{2}$, or explicitly,
\begin{equation}
\label{eq:photon-sphere-radial}
\frac{\mathrm{d}V_{\rm eff}}{\mathrm{d}r}\Big|_{r_{\rm ph}}=0
\quad\Longrightarrow\quad
r_{\rm ph}\,f'(r_{\rm ph}) - 2\,f(r_{\rm ph}) = 0.
\end{equation}

Once $r_{\rm ph}$ is known, evaluating Eq.~\eqref{eq:impact-parameter} at $r_{\rm ph}$ gives the shadow radius seen by a distant static observer,
\begin{equation}
\label{eq:shadow-radius-def}
R_{\rm sh}=b_{\rm ph}=\frac{r_{\rm ph}}{\sqrt{f(r_{\rm ph})}}.
\end{equation}
In the Schwarzschild limit $f(r)=1-2M/r$, Eqs.~\eqref{eq:photon-sphere-radial} and \eqref{eq:shadow-radius-def} give the familiar values
\begin{equation}
r_{\rm ph}^{(\rm Sch)}=3M,\qquad R_{\rm sh}^{(\rm Sch)}=3\sqrt{3}\,M\approx5.196\,M.
\end{equation}
Below we examine how the quantum corrections modify these results for BH-I and BH-II.

\subsection{Photon Sphere of BH-I}\label{subsec:photon-shadow-BH-I}

For BH-I the metric function in terms of $u\equiv M/r$ reads
\begin{equation}
f_{\rm I}(u)=1-2u+\left(\frac{\xi}{M}\right)^{2}u^{2s}(1-2u)^{2},
\end{equation}
and the photon-sphere condition can be expressed as
\begin{equation}
\label{eq:photon-sphere-BH1}
h_{\rm I}(u_{\rm ph})=0,\qquad h_{\rm I}(u)=2-6u+2\left(\frac{\xi}{M}\right)^{2}u^{2s}(1-2u)\bigl[(s+1)-2u(s+2)\bigr],
\end{equation}
in which $u_{\rm ph}=M/r_{\rm ph}$.
In the Schwarzschild limit $\xi=0$ this reduces to $2-6u=0$ and yields $r_{\rm ph}=3M$. With quantum corrections the photon sphere radius $r_{\rm ph}$ is defined as the outermost solution of Eq.~\eqref{eq:photon-sphere-BH1} and acquires a non-trivial $(\xi,s)$ dependence that must be solved numerically in general.

One has $h_{\rm I}(0^{+})=2>0$ and $h_{\rm I}(1/2)=-1<0$, so by the intermediate value theorem a root exists in $(0,1/2)$ for every $(\xi,s)$. To see that this root is unique, denote $p(u)\equiv(s+1)-2u(s+2)$, and write $\eta=(\xi/M)^{2}$. Solving Eq.~\eqref{eq:photon-sphere-BH1} for $\eta$ gives
\begin{equation}
\eta(u_{\rm ph})=\frac{3u_{\rm ph}-1}{u_{\rm ph}^{2s}(1-2u_{\rm ph})\,p(u_{\rm ph})}.
\end{equation}
For all $s>1/2$ and $\eta>0$, the monotonicity of $\eta(u)$ on the relevant interval guarantees a unique $u_{\rm ph}\in(0,1/2)$. At $s=1$, Eq.~\eqref{eq:photon-sphere-BH1} factorizes as $h_{\rm I}(u)=2(1-3u)[1+2\eta u^{2}(1-2u)]$ and the unique root is $u_{\rm ph}=1/3$ regardless of $\eta$. In all cases $u_{\rm ph}<1/2$, so the photon sphere lies outside the horizon; consequently BH-I never exhibits a photon-sphere critical curve.

A remarkable analytical simplification occurs at $s=1$. The photon-sphere condition becomes
\begin{equation}
2(1-3u_{\rm ph})\left[1+2\left(\frac{\xi}{M}\right)^{2}u_{\rm ph}^{2}(1-2u_{\rm ph})\right]=0.
\end{equation}
The second factor satisfies $1+(\xi/M)^{2}u_{\rm ph}^{2}(1-2u_{\rm ph})>0$ for all $u_{\rm ph}\in(0,\frac{1}{2})$ and all $\xi/M$, because $u_{\rm ph}^{2}>0$ and $(1-2u_{\rm ph})>0$ throughout this interval. Therefore the only root comes from the first factor,
\begin{equation}
1-3u_{\rm ph}=0\quad\Longrightarrow\quad u_{\rm ph}=\frac{1}{3},
\end{equation}
i.e.\ $r_{\rm ph}=3M$, for any value of $\xi/M$. The quantum correction enters the photon-sphere condition in precisely such a way that its contribution factorises and cancels identically at $s=1$.

The origin of this cancellation can be traced to the structure of Eq.~\eqref{eq:photon-sphere-BH1}.  At $s=1$ the bracket $p(u)=(s+1)-2u(s+2)$ reduces to $2-6u$, which is exactly the classical photon-sphere function; the quantum term in $h_{\rm I}$ therefore becomes proportional to the classical term, and $h_{\rm I}$ factorises with the universal root $u_{\rm ph}=1/3$.  Physically, with $\mu\propto 1/r$ the correction $\delta f_{\rm I}\propto u^{2}(1-2u)^{2}$ has a logarithmic derivative $u\,\mathrm{d}\ln\delta f_{\rm I}/\mathrm{d}u$ whose radial profile mirrors that of the Schwarzschild mass term under the photon-sphere operator $2+u\,\mathrm{d}/\mathrm{d}u$.  The cancellation is thus an exact algebraic property of the $\bar{\mu}$ polymerisation together with the horizon-vanishing prefactor $(1-2M/r)^{2}$, rather than an accident of parameter tuning; no comparable factorisation occurs for any $s\neq1$ or for BH-II.

This is only a position restoration: $f_{\rm I}(3M)=\frac13+\frac{(\xi/M)^2}{81}>\frac13$, so the shadow radius
\begin{equation}
\label{eq:s1-shadow-radius}
R_{\rm sh}^{(s=1)}=\frac{27M}{\sqrt{27+(\xi/M)^{2}}}<3\sqrt{3}\,M
\end{equation}
remains below the Schwarzschild value for all $\xi>0$. The coordinate $r_{\rm ph}$ is restored, but the physical observable $R_{\rm sh}$ retains a quantum imprint.

\begin{figure}[htbp]
\centering
\includegraphics[width=\textwidth]{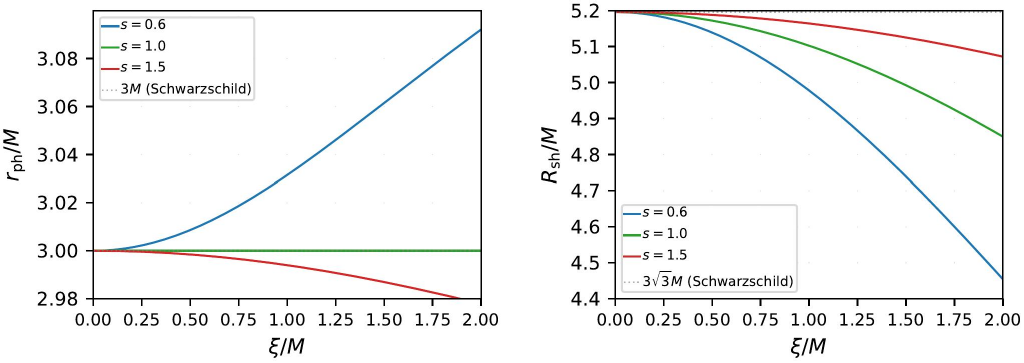}
\caption{BH-I: photon sphere radius $r_{\rm ph}$ (left) and shadow radius $R_{\rm sh}$ (right) as functions of $\xi/M$ for $s=0.6$ (blue), $s=1.0$ (green) and $s=1.5$ (red). The horizontal grey dotted lines mark the Schwarzschild values $r_{\rm ph}^{(\rm Sch)}=3M$ (left panel) and $R_{\rm sh}^{(\rm Sch)}=3\sqrt{3}\,M\approx5.196M$ (right panel).}
\label{fig:BH1-rph-Rsh-xi}
\end{figure}

Figure~\ref{fig:BH1-rph-Rsh-xi} shows $r_{\rm ph}$ (left) and $R_{\rm sh}$ (right) versus $\xi/M$. For $s<1$, $r_{\rm ph}$ increases with $\xi/M$ (outward shift); for $s>1$, it decreases (inward shift); at $s=1$, $r_{\rm ph}=3M$ independent of $\xi$ (exact cancellation). The shadow radius $R_{\rm sh}$ always decreases because $f_{\rm I}(r_{\rm ph})>1/3$; even at $s=1$ where $r_{\rm ph}$ is restored, $R_{\rm sh}$ varies through $f_{\rm I}(3M)$. As follows from the existence proof above ($x_{\rm ph}>2$) and from $R_{\rm sh}=r_{\rm ph}/\sqrt{f_{\rm I}(r_{\rm ph})}$ with $f_{\rm I}(r_{\rm ph})<1$, the three characteristic radii obey the ordering $r_{\rm h}=2M<r_{\rm ph}<R_{\rm sh}$.

Table~\ref{tab:BH1-rph-Rsh-vs-s} lists representative values of $r_{\rm ph}$ and $R_{\rm sh}$ versus $s$. In all cases the $s$-dependence is non-monotonic for $r_{\rm ph}$ (due to the cancellation at $s=1$) but monotonic for $R_{\rm sh}$.

\begin{table}[htbp]
\centering
\caption{BH-I: photon-sphere radius $r_{\rm ph}$ and shadow radius $R_{\rm sh}$ as functions of $s$ for fixed $\xi$. Deviations are relative to Schwarzschild ($r_{\rm ph}^{(\rm Sch)}=3M$, $R_{\rm sh}^{(\rm Sch)}=3\sqrt{3}\,M\approx5.196M$).}
\label{tab:BH1-rph-Rsh-vs-s}
\small
\begin{tabular}{ccccc}
\toprule
& \multicolumn{2}{c}{$\xi=0.3M$} & \multicolumn{2}{c}{$\xi=M$} \\
\cmidrule(lr){2-3}\cmidrule(lr){4-5}
$s$ & $r_{\rm ph}/M$ & $R_{\rm sh}/M$ & $r_{\rm ph}/M$ & $R_{\rm sh}/M$ \\
\midrule
$0.75$ & $3.001$ ($+0.05\%$) & $5.181$ ($-0.3\%$) & $3.014$ ($+0.5\%$) & $5.037$ ($-3.1\%$) \\
$1.00$ & $3.000$ ($0\%$) & $5.188$ ($-0.2\%$) & $3.000$ ($0\%$) & $5.103$ ($-1.8\%$) \\
$1.50$ & $2.999$ ($-0.02\%$) & $5.193$ ($-0.06\%$) & $2.994$ ($-0.2\%$) & $5.164$ ($-0.6\%$) \\
$2.00$ & $3.000$ ($-0.01\%$) & $5.195$ ($-0.02\%$) & $2.996$ ($-0.1\%$) & $5.186$ ($-0.2\%$) \\
\bottomrule
\end{tabular}
\end{table}

\subsection{Photon Sphere of BH-II}\label{subsec:photon-shadow-BH-II}

For BH-II the metric function in terms of $u\equiv M/r$ reads
\begin{equation}
f_{\rm II}(u)=1-2u+4\left(\frac{\xi}{M}\right)^{2}u^{2s+2},
\end{equation}
and the photon-sphere condition becomes
\begin{equation}
\label{eq:photon-sphere-BH2}
h_{\rm II}(u_{\rm ph})=0,\qquad h_{\rm II}(u)=6u-2-8(s+2)\left(\frac{\xi}{M}\right)^{2}u^{2s+2},
\end{equation}
Unlike BH-I, no exact analytical cancellation occurs at any special value of $s$. For fixed $(\xi/M,s)$ the equation must be solved numerically, and a physical photon sphere exists only when the solution satisfies $r_{\rm ph}>r_{\rm h}$ and $f_{\rm II}(r_{\rm ph})>0$. The shadow radius is given by the same expression $R_{\rm sh}=r_{\rm ph}/\sqrt{f_{\rm II}(r_{\rm ph})}$, but now $f_{\rm II}(3M)=1/3+4(\xi/M)^{2}/3^{2s+2}>0$ for all $\xi>0$, so there is no counterpart of the $s=1$ position restoration.

\begin{figure}[htbp]
\centering
\includegraphics[width=0.72\textwidth]{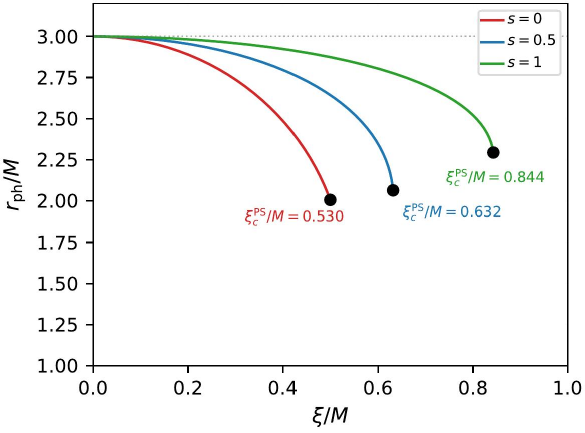}
\caption{BH-II: photon sphere radius $r_{\rm ph}$ vs $\xi/M$ for $s=0$ (red), $s=0.5$ (blue) and $s=1$ (green). Solid dots mark the critical points $(\xi_c^{\rm PS}, r_{\rm ph}^{c})$ where the photon sphere disappears. The grey dotted line marks $r=3M$.}
\label{fig:BH2-rph-xi}
\end{figure}

Figure~\ref{fig:BH2-rph-xi} shows $r_{\rm ph}$ versus $\xi/M$ for three representative $s$ values; each curve starts at the Schwarzschild limit ($3M$) and decreases monotonically as $\xi/M$ increases, and terminates at $(\xi_c^{\rm PS}, r_{\rm ph}^{c})$.

BH-II possesses a photon-sphere critical curve, with the critical amplitude denoted $\xi_c^{\rm PS}$ (PS means photon sphere; the critical radius of the photon sphere itself is denoted $r_{\rm ph}^{c}$ below). The photon-sphere condition reads
\begin{equation}
2-6u_{\rm ph}+(2s+4)\,X=0,\qquad X\equiv4\Bigl(\frac{\xi}{M}\Bigr)^{2}u_{\rm ph}^{2s+2},
\end{equation}
which gives $X=(3u_{\rm ph}-1)/(s+2)$. Treating $\xi/M$ as a function of $u$ via $(\xi/M)^{2}=X/(4u_{\rm ph}^{2s+2})$ and denoting $u_{*}\equiv M/r_{\rm ph}^{c}$, the fold condition $\partial(\xi/M)/\partial u_{\rm ph}=0$ at $u_{*}$ yields
\begin{equation}
\frac{3}{s+2}=\frac{2(s+1)}{u_{*}}\,\frac{3u_{*}-1}{s+2}
\quad\Longrightarrow\quad
u_{*}=\frac{2(s+1)}{3(2s+1)}.
\end{equation}
Substituting back gives $X_{*}=1/[(2s+1)(s+2)]$ and the closed-form critical curve
\begin{equation}
\label{eq:xi-c-PS-BH2}
\frac{\xi_{c}^{\rm PS}(s)}{M}=\frac{3^{s+1}(2s+1)^{s+1/2}}{2^{s+2}(s+1)^{s+1}\sqrt{s+2}}.
\end{equation}
\begin{equation}
\label{eq:rph-crit-BH2}
r_{\rm ph}^{c}=\frac{M}{u_{*}}=\frac{3(2s+1)}{2(s+1)}\,M.
\end{equation}

Interestingly, comparing Eq.~\eqref{eq:rph-crit-BH2} with Eq.~\eqref{eq:rh-c}, we find that $r_{\rm ph}^{c}=\frac{3}{2}\,r_{\rm h}^{c}$.
For $\xi<\xi_{c}^{\rm PS}(s)$ a photon sphere exists outside the horizon; for $\xi>\xi_{c}^{\rm PS}(s)$ it does not. The critical curve gives $\xi_{c}^{\rm PS}(s)/M=3/(4\sqrt{2})\approx0.530$ at $s=0$, $\approx0.632$ at $s=0.5$, and $\approx0.844$ at $s=1$; it rises monotonically with $s$, diverging as $s\to\infty$.

\begin{figure}[htbp]
\centering
\includegraphics[width=0.72\textwidth]{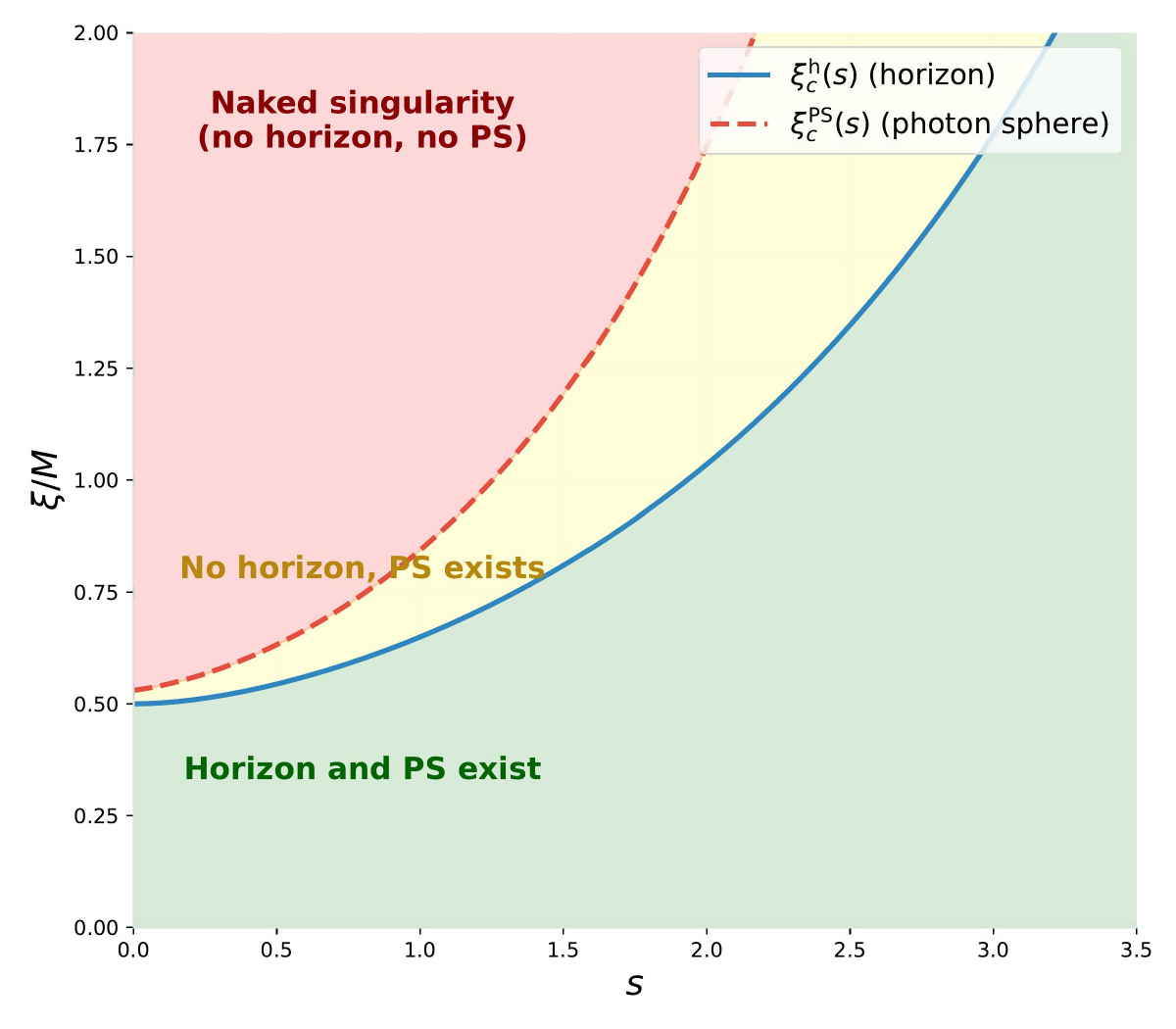}
\caption{BH-II: critical curves in the $(s,\xi/M)$ plane. The solid blue curve is $\xi_c^{\rm h}(s)$ (horizon disappearance); the dashed red curve is $\xi_c^{\rm PS}(s)$ (photon sphere disappearance). Green region (below both curves): horizon and photon sphere both exist. Yellow region (between curves): no horizon, photon sphere exists. Red region (above both curves): naked singularity, no photon sphere.}
\label{fig:BH2-critical-PS}
\end{figure}

Figure~\ref{fig:BH2-critical-PS} compares the two critical curves. For all $s\geq0$ the photon-sphere critical curve lies above the horizon critical curve ($\xi_c^{\rm PS}>\xi_c^{\rm h}$), creating three physically distinct regions in the $(s,\xi/M)$ plane. The green (lower) region: $\xi<\xi_c^{\rm h}(s)$, where both a regular event horizon and a photon sphere exist. The yellow (middle) region: $\xi_c^{\rm h}(s)<\xi<\xi_c^{\rm PS}(s)$, where the event horizon has disappeared but the photon sphere still survives (a naked singularity with a photon sphere). The red (upper) region: $\xi>\xi_c^{\rm PS}(s)$, where neither horizon nor photon sphere exists. In this naked-singularity regime the absence of a photon sphere means no photon trapping, so the characteristic photon-ring substructure disappears entirely. By the cosmic censorship hypothesis such configurations are regarded as unphysical; our analysis therefore restricts to the green region $\xi<\xi_c^{\rm h}(s)$. The separation between $\xi_c^{\rm h}$ and $\xi_c^{\rm PS}$ widens as $s$ increases, from $\Delta\xi/M\approx0.03$ at $s=0$ to $\Delta\xi/M\approx0.19$ at $s=1$.

For BH-II the radii satisfy $r_{\rm h}<r_{\rm ph}<3M<R_{\rm sh}$. As $\xi$ increases, both $r_{\rm h}$ and $r_{\rm ph}$ shift inward, reducing $R_{\rm sh}$ below the Schwarzschild value $3\sqrt{3}\,M$.

Table~\ref{tab:BH2-rph-Rsh-vs-s} gives the $s$-dependence at fixed $\xi$. Note that for $\xi=0.8M$, when $s\leq0.75$ no photon sphere exists because $\xi>\xi_{c}^{\rm PS}(s)$. In all cases both radii increase monotonically with $s$ and approach Schwarzschild from below, with the convergence rate set by the $r^{-(2s+2)}$ falloff of the quantum correction.

\begin{table}[htbp]
\centering
\caption{BH-II: photon-sphere radius $r_{\rm ph}$ and shadow radius $R_{\rm sh}$ as functions of $s$ for fixed $\xi$. Deviations are relative to Schwarzschild. A dash indicates that no photon sphere exists ($\xi>\xi_{c}^{\rm PS}(s)$).}
\label{tab:BH2-rph-Rsh-vs-s}
\small
\begin{tabular}{ccccc}
\toprule
& \multicolumn{2}{c}{$\xi=0.3M$} & \multicolumn{2}{c}{$\xi=0.8M$} \\
\cmidrule(lr){2-3}\cmidrule(lr){4-5}
$s$ & $r_{\rm ph}/M$ & $R_{\rm sh}/M$ & $r_{\rm ph}/M$ & $R_{\rm sh}/M$ \\
\midrule
$0.00$ & $2.737$ ($-8.8\%$) & $4.859$ ($-6.5\%$) & --- & --- \\
$0.25$ & $2.830$ ($-5.7\%$) & $5.003$ ($-3.7\%$) & --- & --- \\
$0.50$ & $2.892$ ($-3.6\%$) & $5.086$ ($-2.1\%$) & --- & --- \\
$0.75$ & $2.933$ ($-2.2\%$) & $5.134$ ($-1.2\%$) & --- & --- \\
$1.00$ & $2.958$ ($-1.4\%$) & $5.160$ ($-0.7\%$) & $2.520$ ($-16.0\%$) & $4.851$ ($-6.6\%$) \\
$1.50$ & $2.984$ ($-0.5\%$) & $5.184$ ($-0.2\%$) & $2.867$ ($-4.4\%$) & $5.103$ ($-1.8\%$) \\
$2.00$ & $2.994$ ($-0.2\%$) & $5.192$ ($-0.1\%$) & $2.955$ ($-1.5\%$) & $5.167$ ($-0.6\%$) \\
\bottomrule
\end{tabular}
\end{table}

Comparing BH-II with BH-I, we find three key distinctions between the two families: (i) BH-I always has a photon sphere for $s>1/2$, while BH-II loses it when $\xi>\xi_{c}^{\rm PS}(s)$; (ii) the $s=1$ exact-cancellation ($r_{\rm ph}=3M$) is unique to BH-I; (iii) BH-II produces larger deviations from Schwarzschild because its correction lacks the $(1-2M/r)^2$ suppression factor.

\subsection{Small-Parameter Expansion of the Photon Sphere Radius}\label{subsec:small-param}

For small quantum corrections $\xi/M\ll 1$, the photon-sphere radius and shadow radius admit systematic expansions around the Schwarzschild values.  The strategy is to expand the photon-sphere condition \eqref{eq:photon-sphere-radial} together with the metric function $f(r)$ to first order in $\eta\equiv(\xi/M)^{2}$.  These expansions provide analytic insight into the $(\xi,s)$ dependence and serve as inputs to the degeneracy analysis presented in Sec.~\ref{subsec:degeneracy}.

For BH-I we write $r_{\rm ph}=3M+\delta r$ with $|\delta r|\ll M$ and substitute into $r f'(r)=2f(r)$.  Expanding the classical part $\bar f(r)=1-2M/r$ and the quantum correction $\delta f_{\rm I}(r)=\xi^{2}r^{-2s}(1-2M/r)^{2}$ about $r=3M$ to first order in $\delta r$ and $\eta$ gives
\begin{align}
f_{\rm I}(3M+\delta r) &= \frac{1}{3} + \frac{2\,\delta r}{9M} + \eta\,3^{-2s-2} + O(\eta^{2}),\\[2pt]
f'_{\rm I}(3M+\delta r) &= \frac{2}{9M} - \frac{4\,\delta r}{27M^{2}} + \frac{2(2-s)}{9}\,\eta\,3^{-2s-1}M^{-1} + O(\eta^{2}).
\end{align}
Inserting these into the photon-sphere condition $r f'(r)=2f(r)$ and collecting the $O(\eta)$ terms yields
\begin{equation}
-\frac{2\,\delta r}{9M} + \frac{2(2-s)}{3}\,\eta\,3^{-2s-1}
= \frac{4\,\delta r}{9M} + \frac{2}{3}\,\eta\,3^{-2s-1} + O(\eta^{2}),
\end{equation}
where the terms proportional to $\delta r$ cancel on the left-hand side because the Schwarzschild photon sphere is an extremum.  Solving for $\delta r$ gives
\begin{equation}
\delta r = (1-s)\,3^{-2s-1}\,\eta\,M + O(\eta^{2}),
\end{equation}
so that
\begin{equation}
\label{eq:rph-expand-BH1}
\frac{r_{\rm ph}^{\rm I}}{M} = 3 + (1-s)\,3^{-2s-1}\,\eta + O(\eta^{2}).
\end{equation}
The coefficient $(1-s)$ makes the exact-cancellation phenomenon manifest: at $s=1$ the $O(\eta)$ term vanishes identically and $r_{\rm ph}=3M$ to all orders in $\xi$.  For $s<1$ the correction is positive ($r_{\rm ph}>3M$), while for $s>1$ it is negative, producing the non-monotonic trajectory seen in Table~\ref{tab:BH1-rph-Rsh-vs-s}.

For BH-II the same procedure gives
\begin{equation}
\label{eq:rph-expand-BH2}
\frac{r_{\rm ph}^{\rm II}}{M} = 3 - 4(s+2)\,3^{-2s-1}\,\eta + O(\eta^{2}),
\end{equation}
where the correction is always negative and grows in magnitude with $s$ through the prefactor $(s+2)$.

Once $r_{\rm ph}$ is known, the shadow radius follows from $R_{\rm sh}=r_{\rm ph}/\sqrt{f(r_{\rm ph})}$.  Expanding $f(r_{\rm ph})$ to $O(\eta)$ and combining with the photon-sphere shift gives
\begin{align}
\label{eq:Rsh-expand-BH1}
\frac{R_{\rm sh}^{\rm I}}{M} &= 3\sqrt{3}\left[1 - \frac{\eta}{2}\,3^{-2s-1} \right] + O(\eta^{2}),\\[2pt]
\label{eq:Rsh-expand-BH2}
\frac{R_{\rm sh}^{\rm II}}{M} &= 3\sqrt{3}\left[1 - 2\,\eta\,3^{-2s-1} \right] + O(\eta^{2}).
\end{align}
Both expansions reproduce the exact numerical results to within $0.5\%$ for $\xi=0.1M$.  Note that even at the exact-cancellation point $s=1$, where $r_{\rm ph}^{\rm I}=3M$ to all orders, the shadow radius retains a quantum imprint because $f_{\rm I}(3M)\neq 1/3$.

The convergence of these expansions is controlled by the dimensionless parameter $\eta=(\xi/M)^{2}$, but the rate of convergence also depends on $s$ through the prefactor $3^{-2s-1}$.  For small $s$ the factor $3^{-2s-1}$ is larger, so the $O(\eta)$ term is more significant and the expansion requires smaller $\eta$ to achieve the same accuracy.  Quantitatively, the expansion is reliable when $\eta\,3^{-2s-1}\ll 1$, or equivalently $\xi/M\ll 3^{s+1/2}$.  At $s=0.5$ this means $\xi/M\ll 3$, while at $s=1$ it means $\xi/M\ll 3^{3/2}\approx 5.2$; the expansion therefore remains accurate well beyond the strict perturbative regime $\xi/M\ll 1$ provided $s$ is not too small.  For very small $s$ (e.g. $s\lesssim 0.1$) the condition tightens and one should compare directly with the exact numerical results.

\subsection{Instability of Photon Spheres and Lyapunov Exponent}

For the photon sphere to produce a bright ring in the observed image, it must correspond to unstable circular null geodesics: any slight perturbation must cause the photon to diverge exponentially from the critical orbit. The instability is quantified by the curvature of the effective potential at $r_{\rm ph}$. Since $V_{\rm eff}'(r_{\rm ph})=0$, the sign of $V_{\rm eff}''(r_{\rm ph})$ determines stability: a local maximum ($V_{\rm eff}''<0$) is unstable, a local minimum ($V_{\rm eff}''>0$) would be stable. For $V_{\rm eff}(r)=f(r)/r^{2}$, the photon-sphere condition yields
\begin{equation}
\label{eq:V-double-prime}
V_{\rm eff}''(r_{\rm ph})=\frac{r_{\rm ph}^{2}f''(r_{\rm ph})-2f(r_{\rm ph})}{r_{\rm ph}^{4}}.
\end{equation}
The Lyapunov exponent, which measures the exponential divergence rate of nearby null geodesics, is
\begin{equation}
\label{eq:lyapunov}
\lambda_{\rm ph}=\sqrt{-\frac{r_{\rm ph}^{2}\,f(r_{\rm ph})\,V_{\rm eff}''(r_{\rm ph})}{2}}=\sqrt{f(r_{\rm ph})\left[\frac{f(r_{\rm ph})}{r_{\rm ph}^{2}}-\frac{f''(r_{\rm ph})}{2}\right]},
\end{equation}
so $\lambda_{\rm ph}$ is real and positive iff $V_{\rm eff}''(r_{\rm ph})<0$.  In geometric units $\lambda_{\rm ph}$ has the dimension of inverse mass; throughout this section $\lambda_{\rm ph}$ is quoted in units of $M^{-1}$ (equivalently, $M\lambda_{\rm ph}$ is dimensionless).

For general $(\xi,s)$ the sign of $V_{\rm eff}''(r_{\rm ph})$ can be established rigorously. The key observation is the identity
\begin{equation}
	\label{eq:Vprime-hI}
	M^{2}\,\frac{\mathrm{d}V_{\rm eff}}{\mathrm{d}u}=u\,h(u),
\end{equation}
which follows by direct differentiation of $V_{\rm eff}=u^{2}f(u)/M^{2}$, with $h_{\rm I}(u)$ and $h_{\rm II}(u)$ defined in Eqs.~\eqref{eq:photon-sphere-BH1} and \eqref{eq:photon-sphere-BH2}. 

For BH-I, on the interval $u\in(0,1/2)$ (i.e.\ outside the horizon, $r>2M$) the metric function $f_{\rm I}(u)=1-2u+\eta\,u^{2s}(1-2u)^{2}$ with $\eta=(\xi/M)^{2}$ is strictly positive, so $V_{\rm eff}>0$ inside the interval while $V_{\rm eff}\to0$ as $u\to0^{+}$ and $V_{\rm eff}(1/2)=0$. By Eq.~\eqref{eq:Vprime-hI} the stationary points of $V_{\rm eff}$ are precisely the roots of $h_{\rm I}(u)=0$. As shown in Sec.~\ref{subsec:photon-shadow-BH-I}, solving $h_{\rm I}(u)=0$ for $\eta$ yields a function that is strictly monotonic on the physical branch: $\eta(u)$ decreases from $+\infty$ to $0$ on $(u_{\rm D},1/3)$ for $s<1$ and increases from $0$ to $+\infty$ on $(1/3,u_{\rm D})$ for $s>1$, where $u_{\rm D}\equiv(s+1)/[2(s+2)]$. Hence for every $\eta>0$ the root $u_{\rm ph}\in(0,1/2)$ is unique. Since $h_{\rm I}(0^{+})=2>0$ and $h_{\rm I}(1/2)=-1<0$, the function $h_{\rm I}$ changes sign from positive to negative at this unique root, that is, $h_{\rm I}(u)>0$ for $u<u_{\rm ph}$ and $h_{\rm I}(u)<0$ for $u>u_{\rm ph}$ on $(0,1/2)$, and in particular $h_{\rm I}'(u_{\rm ph})<0$. The effective potential therefore increases on $(0,u_{\rm ph})$ and decreases on $(u_{\rm ph},1/2)$, so the photon sphere is the unique global maximum of $V_{\rm eff}$. Converting back to the radial coordinate with $\mathrm{d}u/\mathrm{d}r=-u^{2}/M$ and evaluating at the stationary point where $h_{\rm I}(u_{\rm ph})=0$, one obtains
\begin{equation}
	\label{eq:V2-hIprime}
	V_{\rm eff}''(r_{\rm ph})=\frac{u_{\rm ph}^{5}}{M^{4}}\,h_{\rm I}'(u_{\rm ph})<0.
\end{equation}
The photon sphere of BH-I is therefore unstable for all admissible $(\xi/M,s)$, and the Lyapunov exponent in Eq.~\eqref{eq:lyapunov} is real and positive throughout the physical parameter space.
For BH-I, using $u_{\rm ph}=M/r_{\rm ph}$ and eliminating $(\xi/M)^{2}$ via the photon-sphere condition gives
\begin{equation}
\label{eq:V2-BH1-explicit}
V_{\rm eff}''(r_{\rm ph})=\frac{u_{\rm ph}^{4}\,\mathcal{N}^{(\rm I)}(u_{\rm ph},s)}{M^{4}\,(1-2u_{\rm ph})\,\bigl[s+1-2(s+2)u_{\rm ph}\bigr]},
\end{equation}
with
\begin{equation}
\begin{aligned}
\mathcal{N}^{(\rm I)}(u,s)&=-4s(s+1)+2u(14s^{2}+19s+3)\\
&\quad-8u^{2}(8s^{2}+15s+4)+24u^{3}(2s^{2}+5s+2).
\end{aligned}
\end{equation}

For BH-II, the metric function $f_{\rm II}(u)=1-2u+4\eta\,u^{2s+2}$ with $\eta=(\xi/M)^{2}$ is strictly positive on $u\in(0,u_{h})$ where $u_{h}=M/r_{h}<1/2$ is the outer-horizon root, and $f_{\rm II}(u_{h})=0$. By Eq.~\eqref{eq:Vprime-hI} the stationary points of $V_{\rm eff}$ in the physical region are the roots of $h_{\rm II}(u)=0$, where $h_{\rm II}(u)=6u-2-8(s+2)\eta\,u^{2s+2}$. At the left boundary $h_{\rm II}(0^{+})=-2<0$; at the right boundary $u\to u_{h}^{-}$ one has $h_{\rm II}(u)\to 6u_{h}-2$. Using $f_{\rm II}(u_{h})=0$ gives $4\eta\,u_{h}^{2s+2}=2u_{h}-1$, so $h_{\rm II}(u_{h})=(6-8s-16)u_{h}+8s+14=-(8s+10)(1-2u_{h})>0$ because $u_{h}<1/2$. Hence $h_{\rm II}$ changes sign from negative to positive at the unique root $u_{\rm ph}\in(0,u_{h})$, and in particular $h_{\rm II}'(u_{\rm ph})>0$.

Differentiating Eq.~\eqref{eq:Vprime-hI} gives $M^{2}\,\mathrm{d}^{2}V_{\rm eff}/\mathrm{d}u^{2}=h_{\rm II}(u)+u\,h_{\rm II}'(u)$. At the stationary point $h_{\rm II}(u_{\rm ph})=0$, and converting back to the radial coordinate with $\mathrm{d}u/\mathrm{d}r=-u^{2}/M$ yields
\begin{equation}
	\label{eq:V2-hIIprime}
	V_{\rm eff}''(r_{\rm ph})=-\frac{u_{\rm ph}^{5}}{M^{4}}\,h_{\rm II}'(u_{\rm ph})<0.
\end{equation}
The photon sphere of BH-II is therefore unstable for all admissible $(\xi/M,s)$.

For BH-II, eliminating $\eta=(\xi/M)^{2}$ via the photon-sphere condition gives the explicit formula
\begin{equation}
\label{eq:V2-BH2-explicit}
V_{\rm eff}''(r_{\rm ph})=\frac{2u_{\rm ph}^{4}\,\bigl[(6s+3)\,u_{\rm ph}-2(s+1)\bigr]}{M^{4}},
\end{equation}
in which the factor $(6s+3)u_{\rm ph}-2(s+1)$ is a linear, strictly increasing function of $u_{\rm ph}$. Since $u_{\rm ph}<u_{V}\equiv 2(s+1)/[3(2s+1)]$ for all admissible parameters (as shown above), this factor is negative and $V_{\rm eff}''(r_{\rm ph})<0$ directly.

\begin{figure}[htbp]
\centering
\includegraphics[width=\textwidth]{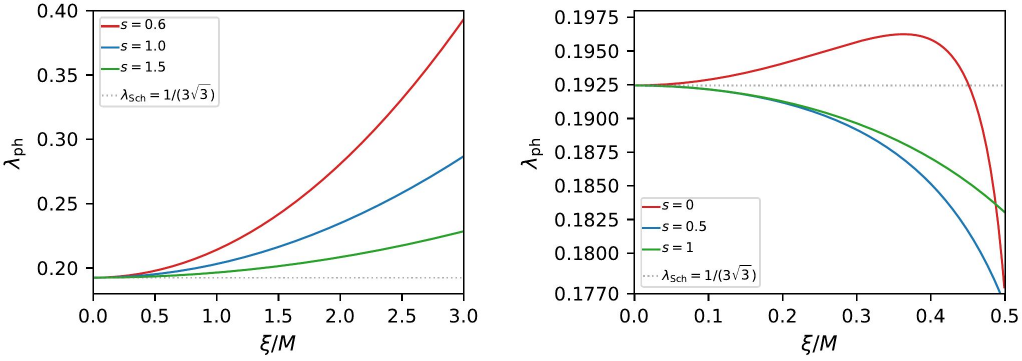}
\caption{Lyapunov exponent $\lambda_{\rm ph}$ vs $\xi/M$. \textbf{(a)} BH-I for $s=0.6$ (red), $s=1.0$ (blue) and $s=1.5$ (green). The grey dotted line marks the Schwarzschild value $1/(3\sqrt{3})\approx0.192$. \textbf{(b)} BH-II for $s=0$ (red), $s=0.5$ (blue) and $s=1$ (green). Each curve terminates at the critical point where the photon sphere disappears.}
\label{fig:lyapunov}
\end{figure}

Figure~\ref{fig:lyapunov} shows the Lyapunov exponent as a function of $\xi/M$ for representative values of $s$. For both metrics $\lambda_{\rm ph}$ deviates from the Schwarzschild value $1/(3\sqrt{3})\approx0.192$, reflecting the modified instability induced by quantum corrections. The deviation is most pronounced at small $s$ where the correction decays slowly. BH-I (left) allows much larger $\xi/M$, while BH-II (right) exhibits the critical-curve phenomenon where each curve terminates at $\xi_c^{\rm PS}(s)$.

For BH-II (Fig.~\ref{fig:lyapunov}(b)) the $s=0$ curve shows a characteristic non-monotonic behavior: it first rises above the Schwarzschild value because the charge-like term $4\xi^{2}/r^{2}$ steepens the potential, then drops as the photon sphere moves inward and the peak flattens. For $s>0$ the faster falloff suppresses this effect and $\lambda_{\rm ph}$ decreases monotonically. Quantitatively, in the Reissner--Nordstr\"om limit the horizon and photon sphere coalesce at $\xi_c^{\rm h}(0)=1/2$ while the photon sphere survives until $\xi_c^{\rm PS}(0)\approx0.530$; the interval $0.500<\xi/M<0.530$ is the yellow region of Fig.~\ref{fig:BH2-critical-PS}.

\subsection{EHT Constraints and Shadow Degeneracy}\label{subsec:eht-constraints}

\subsubsection{EHT Observational Constraints}

The shadow angular diameter of a black hole at distance $D$ is
\begin{equation}
\label{eq:shadow-angle}
\theta_{\rm sh} = 2\arcsin\left(\frac{R_{\rm sh}}{D}\right) \approx \frac{2R_{\rm sh}}{D},
\end{equation}
where the last equality holds because $R_{\rm sh}/D\ll 1$ for both M87* and Sgr~A*. Since $R_{\rm sh}=r_{\rm ph}/\sqrt{f(r_{\rm ph})}$, any departure of $(\xi,s)$ from the Schwarzschild values shifts $\theta_{\rm sh}$ away from the Einstein-predicted angular size. The EHT has measured
\begin{equation}
\theta_{\rm sh}^{(\mathrm{M87*})} = 42 \pm 3\;\mu\mathrm{as}, \qquad \theta_{\rm sh}^{(\mathrm{Sgr\,A*})} = 48.7 \pm 7\;\mu\mathrm{as},
\end{equation}
providing two independent windows on the $(\xi,s)$ parameter space.

We first consider the constraints on BH-I. The $s=1$ exact cancellation makes the shadow radius relatively insensitive to $\xi$, while for $s\neq 1$ the deviation from Schwarzschild grows with $\xi/M$. The resulting bounds are collected in Table~\ref{tab:eht-BH1}.

\begin{table}[htbp]
\centering
\caption{EHT constraints on BH-I: maximum permitted $\xi/M$ from M87* ($\sim 7\%$ $1\sigma$ diameter uncertainty) and Sgr~A* ($\sim 14\%$ $1\sigma$) shadow-diameter measurements at representative $s$. The $s=1$ column shows the weakest constraint due to the exact $r_{\rm ph}=3M$ cancellation.}
\label{tab:eht-BH1}
\begin{tabular}{@{}lcccc@{}}
\toprule
& $s=0.55$ & $s=1$ & $s=1.5$ & $s=2$ \\
\midrule
Shadow deviation at $\xi/M=0.3$ & $\sim 0.4\%$ & $\sim 0.1\%$ & $\sim 0.1\%$ & $\sim 0.0\%$ \\
Shadow deviation at $\xi/M=1.0$ & $\sim 4.7\%$ & $\sim 1.8\%$ & $\sim 0.6\%$ & $\sim 0.2\%$ \\
\midrule
M87* constraint ($\xi\lesssim$) & $1.2M$ & $2.1M$ & $3.4M$ & $5.8M$ \\
Sgr~A* constraint ($\xi\lesssim$) & $1.7M$ & $3.1M$ & $4.8M$ & $8.3M$ \\
\bottomrule
\end{tabular}
\end{table}

\begin{table}[htbp]
\centering
\caption{EHT constraints on BH-II (same protocol as Table~\ref{tab:eht-BH1}). Dashes indicate that no regular horizon exists for that parameter combination ($\xi>\xi_c^{\rm h}(s)$). Entries marked $^{*}$ are limited by the horizon-existence bound rather than by the EHT measurement (the two coincide at $s=1$, where $\xi_c^{\rm h}(1)\approx0.65M$).}
\label{tab:eht-BH2}
\begin{tabular}{@{}lcccc@{}}
\toprule
& $s=0$ & $s=0.5$ & $s=1$ & $s=2$ \\
\midrule
Shadow deviation at $\xi/M=0.3$ & $\sim 6.5\%$ & $\sim 2\%$ & $\sim 0.7\%$ & $\sim 0.1\%$ \\
Shadow deviation at $\xi/M=0.6$ & --- & --- & $\sim 3\%$ & $\sim 0.3\%$ \\
\midrule
M87* constraint ($\xi\lesssim$) & $0.3M$ & $0.5M$ & $0.65M^{*}$ & $1.0M$ \\
Sgr~A* constraint ($\xi\lesssim$) & $0.4M$ & $0.5M$ & $0.65M^{*}$ & $1.0M$ \\
\bottomrule
\end{tabular}
\end{table}

Two important caveats apply to these constraints. The quoted bounds constrain the combination $\xi=\zeta_{0}b^{s}/a$ (not the bare polymerisation scale $\zeta_{0}$), so a larger auxiliary coefficient $a$ weakens the bound. In addition, EHT measurements carry systematics beyond the quoted statistical errors, most notably the mass-to-distance ratio uncertainty; moreover, for optically thin accretion flows the observed ring diameter can be offset from the geometric shadow boundary $R_{\rm sh}$ by emission-profile effects~\cite{Gralla2019}. The bounds in Tables~\ref{tab:eht-BH1} and~\ref{tab:eht-BH2} should therefore be regarded as order-of-magnitude estimates.

Operationally, each bound is obtained by requiring the fractional shadow-radius deviation $|\delta|\equiv|R_{\rm sh}-R_{\rm sh}^{\rm Schw}|/R_{\rm sh}^{\rm Schw}$ to remain below the $1\sigma$ fractional uncertainty of the measured shadow diameter ($\sim7\%$ for M87*, $\sim14\%$ for Sgr~A*), with the source mass-to-distance ratio fixed at its fiducial value. A statistically rigorous treatment would marginalise over the mass-to-distance priors and the astrophysical systematics of the image reconstruction; such a Bayesian analysis is beyond the scope of the present work, and the bounds quoted here should be read as illustrative rather than robust exclusion limits.

Comparing the two families reveals an important observational discriminant. For comparable $(\xi,s)$, BH-II is constrained roughly $2$--$4$ times more tightly than BH-I because its correction term is not suppressed by $(1-2M/r)^2$. The two families are therefore distinguished not only by their theoretical construction but also by their observational signatures: a future measurement of $R_{\rm sh}$ at $\sim 1\%$ precision could exclude large regions of the BH-II parameter space while leaving BH-I essentially unconstrained. The $s$-dependence is also qualitatively different: BH-I constraints are weakest at $s=1$ (exact cancellation), whereas BH-II constraints are strongest at small $s$ (slow decay) and weaken monotonically with $s$.

A complementary class of weak-field constraints could also arise: for BH-I with $1/2<s<1$ the correction decays as $(M/r)^{2s}$, more slowly than the second post-Newtonian term, and for BH-II at $s=0$ as $1/r^{2}$, so precision tracking of the S-stars around Sgr~A* ($r\sim10^{3}$--$10^{4}\,M$) could bound $\xi/M$ in the slow-decay regime independently of the shadow size. Such constraints are expected to be most competitive precisely where the shadow deviations are also largest; a detailed analysis is left to future work.

Looking ahead, space-based very-long-baseline interferometry (VLBI), such as the Black Hole Explorer (BHEX), could reach $\sim 1\%$ shadow-diameter precision, tightening the BH-II bound to $\xi\lesssim 0.1M$ at $s=1$. A simultaneous measurement of the Lyapunov exponent $\lambda_{\rm ph}$, which carries a different $(\xi,s)$ dependence from $R_{\rm sh}$, would break the shadow degeneracy and provide a powerful cross-check on quantum corrections in the strong-field regime.


\subsubsection{Shadow Degeneracy and Parameter Breaking}\label{subsec:degeneracy}

\begin{figure}[htbp]
\centering
\includegraphics[width=\textwidth]{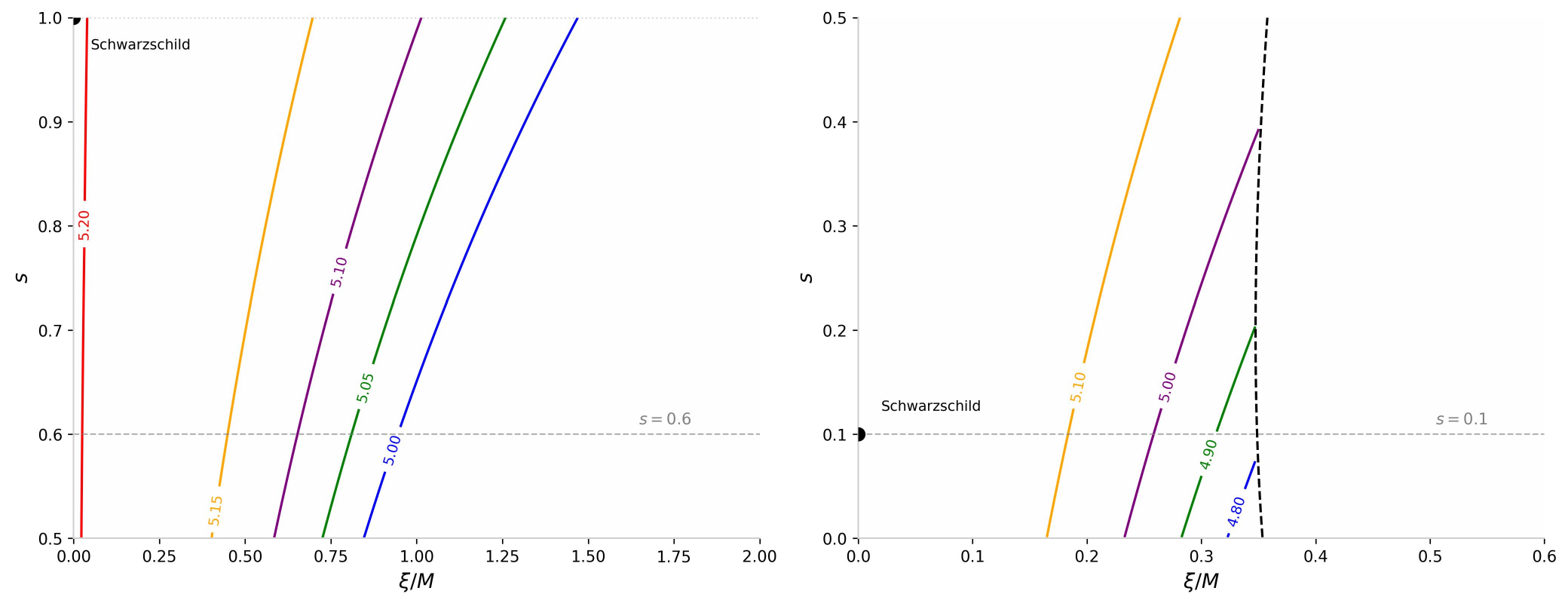}
\caption{Constant shadow-radius contours in the $(\xi,s)$ parameter plane.  \textbf{Left:} BH-I, with $s$ restricted to $(0.5,1.0]$ (where the ADM mass is well defined) and the reference line $s=0.6$ marked by a horizontal dashed grey line; the Schwarzschild point $(\xi=0,s=1)$ is marked by a black dot.  \textbf{Right:} BH-II, with $s$ restricted to $[0,0.5]$ and the reference line $s=0.1$ marked similarly; the photon-sphere critical curve (black dashed) bounds the region where a photon sphere exists.  In both metrics $R_{\rm sh}$ decreases monotonically with $\xi$ from the Schwarzschild value $3\sqrt{3}\,M$, so contours with smaller $R_{\rm sh}$ lie at larger $\xi$.}
\label{fig:degeneracy}
\end{figure}

Since EHT measurements constrain a single angular scale $\theta_{\rm sh}\propto R_{\rm sh}/D$ while the theory contains two parameters $(\xi,s)$, a single shadow measurement inevitably suffers an observational degeneracy: many different $(\xi,s)$ pairs can produce the same $R_{\rm sh}$.  Figure~\ref{fig:degeneracy} shows the constant-$R_{\rm sh}$ contours in the $(\xi,s)$ plane for both metric families.

For BH-I the degeneracy is particularly striking near $s=0.6$: a measurement of $R_{\rm sh}\approx 5.19M$ could correspond to either the Schwarzschild point $(\xi=0)$ or a quantum-corrected point with $(\xi\approx 0.16M,s=0.6)$.  The exact-cancellation locus $s=1$ adds another layer: there $r_{\rm ph}=3M$ for any $\xi$, yet $R_{\rm sh}$ still varies through $f(r_{\rm ph})\neq 1/3$.  For BH-II at $s=0.1$ the situation is similar but milder: both $r_{\rm ph}$ and $R_{\rm sh}$ decrease monotonically with $\xi$, but a single $R_{\rm sh}$ measurement still leaves a one-parameter family of compatible $(\xi,s)$ values (Fig.~\ref{fig:degeneracy}, right).

The degeneracy can be broken by an additional observable.  The Lyapunov exponent $\lambda_{\rm ph}$ provides a natural discriminant because it depends on $(\xi,s)$ through a different combination than $R_{\rm sh}$ does.  Expanding Eq.~\eqref{eq:lyapunov},
\begin{equation}
\lambda_{\rm ph}^{2} = f(r_{\rm ph})\Bigl[\frac{f(r_{\rm ph})}{r_{\rm ph}^{2}} - \frac{f''(r_{\rm ph})}{2}\Bigr],
\end{equation}
to $O(\eta)$ with $r_{\rm ph}=3M+\delta r$ and $\eta\equiv(\xi/M)^{2}$, one finds that the shift $\delta r$ drops out (reflecting the extremal nature of $r_{\rm ph}$); only the metric correction $\delta f$ and its second derivative at $r=3M$ survive, giving
\begin{equation}
\lambda_{\rm ph} = \frac{1}{3\sqrt{3}} + c_{\lambda}(s)\,\eta + O(\eta^{2}),
\end{equation}
with the $s$-dependent coefficients
\begin{align}
\label{eq:clambda-BH1}
c_{\lambda}^{\rm I}(s) &= \frac{\sqrt{3}}{18}\,3^{-2s-2}\bigl(4+7s-2s^{2}\bigr),\\[2pt]
\label{eq:clambda-BH2}
c_{\lambda}^{\rm II}(s) &= \frac{2\sqrt{3}}{9}\,3^{-2s-2}\bigl(1-5s-2s^{2}\bigr).
\end{align}

Several features are noteworthy. For BH-I the exact-cancellation locus $s=1$ does not restore $\lambda_{\rm ph}$ to its Schwarzschild value; instead $c_{\lambda}^{\rm I}(1)=\sqrt{3}/162>0$, so the photon sphere becomes more unstable, mirroring the behaviour of $R_{\rm sh}$. For BH-II, $c_{\lambda}^{\rm II}(s)$ changes sign at $s\approx 0.19$: for very small $s$ the Lyapunov exponent is enhanced, while for $s\gtrsim 0.2$ it is suppressed. Thus $\lambda_{\rm ph}$ carries a genuinely different $(\xi,s)$ fingerprint from $R_{\rm sh}$, and a joint measurement of both observables would over-determine the parameter space and lift the degeneracy. Note that although $\lambda_{\rm ph}$ is not directly imaged, it is imprinted on the photon-ring substructure: each successive subring is demagnified and dimmer than the previous one by a constant flux ratio controlled by the Lyapunov exponent, which is about $1/23$ in the Schwarzschild limit. Resolving this subring ladder with BHEX or the next-generation EHT would therefore furnish the second observable needed to lift the $(\xi,s)$ degeneracy.

\section{Discussion}\label{sec:discussion}

We have presented a comprehensive analysis of photon sphere and shadow observables for the BH-I and BH-II metric families within the general $\mu$-scheme. Several key results emerge from our analysis.

For BH-I, the photon sphere exists for all $\xi>0$ and $s>1/2$, with the outer horizon fixed at $r=2M$. The exact cancellation at $s=1$ restores $r_{\rm ph}=3M$ for any $\xi$, but the shadow radius retains a quantum imprint because $f_{\rm I}(3M)$ exceeds the Schwarzschild value. The small-parameter expansions derived in Sec.~\ref{sec:shadows} make this structure explicit: the photon-sphere shift vanishes at $s=1$ while the shadow-radius coefficient $-3^{-2s-1}/2$ does not, and the Lyapunov-exponent coefficient $c_{\lambda}^{\rm I}(1)=\sqrt{3}/162$ is positive, so the photon sphere becomes more unstable at the cancellation locus.

For BH-II, both the horizon and the photon sphere exhibit critical curves in the $(\xi,s)$ parameter plane. The photon sphere survives longer than the horizon ($\xi_c^{\rm PS} > \xi_c^{h}$ for all $s$), but disappears when $\xi$ exceeds the closed-form critical value given in Eq.~\eqref{eq:xi-c-PS-BH2}. For comparable parameters, BH-II typically produces larger deviations from Schwarzschild than BH-I because the correction term is not suppressed by the $(1-2M/r)^2$ factor. Our systematic scans (Tables~\ref{tab:BH1-rph-Rsh-vs-s}--\ref{tab:BH2-rph-Rsh-vs-s}) reveal that for both families the shadow radius converges monotonically toward the Schwarzschild value as $s$ increases, with BH-I showing a characteristic non-monotonic $r_{\rm ph}$ trajectory due to the $s=1$ exact cancellation.

We proved analytically that the photon sphere is always unstable ($V''_{\rm eff}(r_{\rm ph})<0$) throughout the physical parameter space of both metrics, supporting the existence of the unstable critical null orbit associated with photon-ring substructure in suitable emission models. The Lyapunov exponent deviates from the Schwarzschild value in a parameter-dependent manner, typically enhanced for BH-I at small $s$ and suppressed for BH-II.

The shadow degeneracy analysis in Sec.~\ref{sec:shadows} shows that a single EHT measurement of $R_{\rm sh}$ cannot uniquely determine both $(\xi,s)$. For BH-I at $s=0.6$, a shadow radius of $R_{\rm sh}\approx 5.19M$ is compatible with both the Schwarzschild point and a quantum-corrected point with $\xi\approx 0.16M$. The degeneracy can be broken by a second observable such as the Lyapunov exponent $\lambda_{\rm ph}$, which carries a different $(\xi,s)$ fingerprint through the coefficients in Eqs.~\eqref{eq:clambda-BH1}--\eqref{eq:clambda-BH2}.

Using EHT measurements of the shadow angular diameter for M87* and Sgr~A*, we derived detailed constraints on the $(\xi,s)$ parameter space for both BH families (Sec.~\ref{subsec:eht-constraints}). M87* provides tighter constraints due to its smaller fractional uncertainty ($\sim 7\%$ vs.~$\sim 14\%$), and BH-II is constrained roughly $2$--$4$ times more tightly than BH-I at comparable $(\xi,s)$ because its correction term lacks the $(1-2M/r)^2$ suppression factor. The two families are also distinguished by their $s$-dependence: BH-I constraints are weakest at $s=1$ (exact cancellation) whereas BH-II constraints are strongest at small $s$ (slow decay).  For convenience, Table~\ref{tab:summary} collects the main analytical results side by side.

\begin{table}[htbp]
\centering
\caption{Summary of the main analytical results for the two metric families. Here $\eta\equiv(\xi/M)^{2}$, and $u_{\rm ph}=M/r_{\rm ph}$ solves Eqs.~\eqref{eq:photon-sphere-BH1} and \eqref{eq:photon-sphere-BH2} respectively.}
\label{tab:summary}
\begin{tabular}{@{}lcc@{}}
\toprule
& BH-I & BH-II \\
\midrule
Quantum correction $\delta f$ & $\eta\,u^{2s}(1-2u)^{2}$ & $4\eta\,u^{2s+2}$ \\
Allowed $s$ range & $s>1/2$ & $s\geq0$ \\
Outer horizon & $2M$ (fixed) & shrinks with $\xi$ \\
Horizon critical curve & none & $\xi_{c}^{\rm h}(s)$, Eq.~\eqref{eq:xi_c_h} \\
PS critical curve & none & $\xi_{c}^{\rm PS}(s)>\xi_{c}^{\rm h}(s)$ \\
Exact result at $s=1$ & $r_{\rm ph}=3M$ for all $\xi$ & --- \\
$r_{\rm ph}$ vs.\ $\xi$ ($s\lesseqqgtr1$) & outward / flat / inward & always inward \\
$R_{\rm sh}$ expansion & $3\sqrt{3}\bigl[1-\tfrac{\eta}{2}3^{-2s-1}\bigr]$ & $3\sqrt{3}\bigl[1-2\eta\,3^{-2s-1}\bigr]$ \\
$\lambda_{\rm ph}$ coefficient $c_{\lambda}(s)$ & $\tfrac{\sqrt{3}}{18}3^{-2s-2}(4+7s-2s^{2})$ & $\tfrac{2\sqrt{3}}{9}3^{-2s-2}(1-5s-2s^{2})$ \\
PS instability & $V_{\rm eff}''<0$ for all $(\xi,s)$ & $V_{\rm eff}''<0$ for all $(\xi,s)$ \\
EHT bound strength & weakest at $s=1$ & strongest at small $s$ \\
\bottomrule
\end{tabular}
\end{table}

Several directions merit further investigation. Ringdown frequencies and gravitational waveforms probe near-horizon geometry where corrections are most pronounced~\cite{ChenYang2025b}, while time-like observables---the innermost stable circular orbit (ISCO), which sets the inner edge of the accretion disk and the maximum frequency of quasi-periodic oscillations (QPOs), and the marginally bound orbit (MBO), which controls the fallback rate in tidal disruption events---carry different $(\xi,s)$ fingerprints from the photon sphere. Extending this analysis to rotating metrics would further strengthen the astrophysical relevance of the framework.

\begin{acknowledgments}
\sloppy
Yu Han is supported by key scientific research projects in universities of Henan Province (Grant No.~25A140014) and Natural Science Foundation of Henan Province (Grant No.~262300421865) and Nanhu Scholars Program for Young Scholars of Xinyang Normal University.
\end{acknowledgments}

\appendix

\begin{center}
\large\textbf{Appendix: Derivation of the covariant quantum Oppenheimer--Snyder model in the general $\mu$-scheme}
\end{center}

\vspace{-0.5em}
\noindent
In this appendix, we derive the metric following the construction in Ref.~\cite{Zhang2022LQC}. Consider an effective mass function $\mathcal{M}(r,k,\mathfrak{u})$ defined by
\begin{equation}\label{eq:EqOS}
\mathcal{M}(r,k,\mathfrak{u}) = \frac{r}{2}
\left[\Bigl(1-\mathfrak{u}^2\Bigr)\,e^{\,i 2\mu(r)\,k}
+ \frac{1}{\mu(r)^2}\,\sin^2\!\bigl(\mu(r)\,k\bigr)\right] ,
\end{equation}
where $k$ is the extrinsic curvature, $\mathfrak{u}$ is the spin-connection variable (distinct from the inverse-radius coordinate $u=M/r$ used in Sec.~\ref{sec:shadows}), and $\mu(r)$ is an arbitrary function of $r$ which vanishes in the classical limit.
The covariant metric reads
\begin{equation}\label{eq:metric-ansatz}
\mathrm{d}s^2 = -f(r,\mathcal{M})\,\mathrm{d}t^2 + \frac{1}{g(r,\mathcal{M})}\,\mathrm{d}r^2 + r^2\,\mathrm{d}\Omega^2 .
\end{equation}
The function $f$ is obtained from $\mathcal{M}$ through
\begin{equation}
\biggl(\frac{\partial \mathcal{M}}{\partial k}\biggr)^2
= r^2\,\frac{g}{f}\,\bigl(\mathfrak{u}^2 - f\bigr) .
\label{eq:f-definition}
\end{equation}
A direct computation from Eq.~\eqref{eq:EqOS} gives $(\partial\mathcal{M}/\partial k)^2$, and using Eq.~\eqref{eq:EqOS} itself to eliminate the complex exponential $e^{\,i2\mu k}$ in favour of $\mathcal{M}$ yields
\begin{equation}
\biggl(\frac{\partial\mathcal{M}}{\partial k}\biggr)^2
= r^2\left(\mathfrak{u}^2 - 1 + \frac{2\mathcal{M}}{r}
- \frac{4\mu^2\mathcal{M}^2}{r^2}\right) .
\label{eq:dM2-simplified}
\end{equation}
Inserting this result into Eq.~\eqref{eq:f-definition}, one finds that the factor $\mathfrak{u}^2-f$ cancels identically and the equation is solved by $f=g=1-2\mathcal{M}/r+4\mu^2(r)\mathcal{M}^2/r^2$. Note that the effective mass function~\eqref{eq:EqOS} contains the complex exponential $e^{\,i2\mu k}$; $\mathcal{M}$ can be real as long as $k$ is allowed to take complex values.
It can be proved that $\mathcal{M}$ is a constant of motion; to reproduce the Schwarzschild metric in the classical limit its value should equal the black hole mass $M$. Substituting the parametrisation of $\mu(r)$ in the general $\mu$-scheme immediately yields the metric components in Eq.~\eqref{eq:BH2_f}.


\begin{thebibliography}{99}

\bibitem{Abbott2016}
B.~P. Abbott \textit{et al.} (LIGO Scientific Collaboration and Virgo Collaboration).
\newblock Observation of gravitational waves from a binary black hole merger.
\newblock \emph{Phys.\ Rev.\ Lett.}, 116:061102, 2016.
\newblock \href{https://doi.org/10.1103/PhysRevLett.116.061102}{doi:10.1103/PhysRevLett.116.061102}, \href{https://arxiv.org/abs/1602.03837}{arXiv:1602.03837}.

\bibitem{EHT2019M87}
{The Event Horizon Telescope Collaboration}.
\newblock First {M87} event horizon telescope results. {I}. {T}he shadow of the supermassive black hole.
\newblock \emph{Astrophys. J. Lett.}, 875:L1, 2019.
\newblock \href{https://doi.org/10.3847/2041-8213/ab0ec7}{doi:10.3847/2041-8213/ab0ec7}.
\newblock \href{https://arxiv.org/abs/1906.11238}{arXiv:1906.11238},

\bibitem{EHT2022SgrA}
{The Event Horizon Telescope Collaboration}.
\newblock First {Sagittarius} {A}* event horizon telescope results. {I}. {T}he shadow of the supermassive black hole in the center of the {M}ilky {W}ay.
\newblock \emph{Astrophys. J. Lett.}, 930:L12, 2022.
\newblock \href{https://doi.org/10.3847/2041-8213/ac6674}{doi:10.3847/2041-8213/ac6674}.
\newblock \href{https://arxiv.org/abs/2311.08680}{arXiv:2311.08680},

\bibitem{Penrose1965}
Roger Penrose.
\newblock Gravitational collapse and space-time singularities.
\newblock \emph{Phys. Rev. Lett.}, 14:57–59, 1965.
\newblock \href{https://doi.org/10.1103/PhysRevLett.14.57}{doi:10.1103/PhysRevLett.14.57}.

\bibitem{Ashtekar2004}
Abhay Ashtekar and Martin Bojowald.
\newblock Quantum geometry and the {S}chwarzschild singularity.
\newblock \emph{Class. Quantum Grav.}, 23:391--411, 2006.
\newblock \href{https://doi.org/10.1088/0264-9381/23/2/008}{doi:10.1088/0264-9381/23/2/008}, \href{https://arxiv.org/abs/gr-qc/0509075}{arXiv:gr-qc/0509075}.

\bibitem{Rovelli2004}
C.~Rovelli.
\newblock \emph{Quantum Gravity}.
\newblock Cambridge University Press, Cambridge, 2004.
\href{https://doi.org/10.1017/CBO9780511755804}{\nolinkurl{doi:10.1017/CBO9780511755804}}.

\bibitem{Ashtekar2006a}
Abhay Ashtekar, Tomasz Pawlowski, and Parampreet Singh.
\newblock Quantum nature of the big bang.
\newblock \emph{Phys. Rev. Lett.}, 96:141301, 2006.
\newblock \href{https://doi.org/10.1103/PhysRevLett.96.141301}{doi:10.1103/PhysRevLett.96.141301}, \href{https://arxiv.org/abs/gr-qc/0602086}{arXiv:gr-qc/0602086}.

\bibitem{Gambini2013}
Rodolfo Gambini and Jorge Pullin.
\newblock Loop quantization of the {S}chwarzschild black hole.
\newblock \emph{Phys. Rev. Lett.}, 110:211301, 2013.
\newblock \href{https://doi.org/10.1103/PhysRevLett.110.211301}{doi:10.1103/PhysRevLett.110.211301}, \href{https://arxiv.org/abs/1302.5265}{arXiv:1302.5265}.

\bibitem{Ashtekar2018}
Abhay Ashtekar, Javier Olmedo, and Parampreet Singh.
\newblock Quantum extension of the Kruskal spacetime.
\newblock \emph{Phys. Rev. D}, 98:126003, 2018.
\newblock \href{https://doi.org/10.1103/PhysRevD.98.126003}{doi:10.1103/PhysRevD.98.126003}, \href{https://arxiv.org/abs/1806.02406}{arXiv:1806.02406}.

\bibitem{Bodendorfer2019}
Norbert Bodendorfer, Fabio M. Mele, and Johannes M{\"u}nch.
\newblock Effective quantum extended spacetime of polymer Schwarzschild black hole.
\newblock \emph{Class. Quant. Grav.}, 36:195015, 2019.
\newblock \href{https://doi.org/10.1088/1361-6382/ab3f16}{doi:10.1088/1361-6382/ab3f16}, \href{https://arxiv.org/abs/1902.04542}{arXiv:1902.04542}.

\bibitem{Bodendorfer2021}
Norbert Bodendorfer, Fabio M. Mele, and Johannes M{\"u}nch.
\newblock (b,v)-type variables for black to white hole transitions in effective loop quantum gravity.
\newblock \emph{Phys. Lett. B}, 819:136390, 2021.
\newblock \href{https://doi.org/10.1016/j.physletb.2021.136390}{doi:10.1016/j.physletb.2021.136390}, \href{https://arxiv.org/abs/1911.12646}{arXiv:1911.12646}.

\bibitem{Bojowald2020}
Martin Bojowald.
\newblock Black-hole models in loop quantum gravity.
\newblock \emph{Universe}, 6:125, 2020.
\newblock \href{https://doi.org/10.3390/universe6080125}{doi:10.3390/universe6080125}, \href{https://arxiv.org/abs/2009.13565}{arXiv:2009.13565}.

\bibitem{Husain2022}
Viqar Husain, J.~G. Kelly, R. Santacruz, and E. Wilson-Ewing.
\newblock Quantum gravity of dust collapse: shock waves from black holes.
\newblock \emph{Phys.\ Rev.\ Lett.}, 128:121301, 2022.
\newblock \href{https://doi.org/10.1103/PhysRevLett.128.121301}{doi:10.1103/PhysRevLett.128.121301}, \href{https://arxiv.org/abs/2109.08667}{arXiv:2109.08667}.

\bibitem{Bojowald2015Covariance}
Martin Bojowald, Suddhasattwa Brahma, and Juan D. Reyes.
\newblock Covariance in models of loop quantum gravity: spherical symmetry.
\newblock \emph{Phys. Rev. D}, 92:045043, 2015.
\newblock \href{https://doi.org/10.1103/PhysRevD.92.045043}{doi:10.1103/PhysRevD.92.045043}.
\newblock \href{https://arxiv.org/abs/1507.00329}{arXiv:1507.00329},

\bibitem{Brahma2018}
Jibril Ben Achour and Suddhasattwa Brahma.
\newblock Covariance in self-dual inhomogeneous models of effective quantum geometry: spherical symmetry and Gowdy systems.
\newblock \emph{Phys. Rev. D}, 97:126003, 2018.
\newblock \href{https://doi.org/10.1103/PhysRevD.97.126003}{doi:10.1103/PhysRevD.97.126003}, \href{https://arxiv.org/abs/1712.03677}{arXiv:1712.03677}.

\bibitem{Gambini2022}
Rodolfo Gambini, Javier Olmedo, and Jorge Pullin.
\newblock Towards a quantum notion of covariance in spherically symmetric loop quantum gravity.
\newblock \emph{Phys.\ Rev.\ D}, 105:026017, 2022.
\newblock \href{https://doi.org/10.1103/PhysRevD.105.026017}{doi:10.1103/PhysRevD.105.026017}, \href{https://arxiv.org/abs/2201.01616}{arXiv:2201.01616}.

\bibitem{Bojowald2022}
Martin Bojowald.
\newblock Comment on ``{T}owards a quantum notion of covariance in spherically symmetric loop quantum gravity''.
\newblock \emph{Phys.\ Rev.\ D}, 105:108901, 2022.
\newblock \href{https://doi.org/10.1103/PhysRevD.105.108901}{doi:10.1103/PhysRevD.105.108901}, \href{https://arxiv.org/abs/2203.06049}{arXiv:2203.06049}.

\bibitem{AlonsoBardaji2022}
Asier Alonso-Bardaji, David Brizuela, and Ra{\"u}l Vera.
\newblock Nonsingular spherically symmetric black-hole model with holonomy corrections.
\newblock \emph{Phys. Rev. D}, 106:024035, 2022.
\newblock \href{https://doi.org/10.1103/PhysRevD.106.024035}{doi:10.1103/PhysRevD.106.024035}, \href{https://arxiv.org/abs/2205.02098}{arXiv:2205.02098}.

\bibitem{AlonsoBardaji2022b}
Asier Alonso-Bardaji, David Brizuela, and Ra{\"u}l Vera.
\newblock An effective model for the quantum {S}chwarzschild black hole.
\newblock \emph{Phys. Lett. B}, 829:137075, 2022.
\newblock \href{https://doi.org/10.1016/j.physletb.2022.137075}{doi:10.1016/j.physletb.2022.137075}, \href{https://arxiv.org/abs/2112.12110}{arXiv:2112.12110}.

\bibitem{AlonsoBardaji2024}
A.~Alonso-Bardaji and D.~Brizuela.
\newblock Spacetime geometry from canonical spherical gravity.
\newblock \emph{Phys. Rev. D}, 109:044065, 2024.
\newblock \href{https://doi.org/10.1103/PhysRevD.109.044065}{doi:10.1103/PhysRevD.109.044065}.
\newblock \href{https://arxiv.org/abs/2310.12951}{arXiv:2310.12951},

\bibitem{Han2024}
Muxin Han and Hongguang Liu.
\newblock Covariant $\bar{\mu}$-scheme effective dynamics, mimetic gravity, and nonsingular black holes: {A}pplications to spherically symmetric quantum gravity.
\newblock \emph{Phys.\ Rev.\ D}, 109:084033, 2024.
\newblock \href{https://doi.org/10.1103/PhysRevD.109.084033}{doi:10.1103/PhysRevD.109.084033}, \href{https://arxiv.org/abs/2212.04605}{arXiv:2212.04605}.

\bibitem{Zhang2025LQLett}
Cong Zhang, Jerzy Lewandowski, Yongge Ma, and Jinsong Yang.
\newblock Black holes and covariance in effective quantum gravity.
\newblock \emph{Phys. Rev. D}, 111:L081504, 2025.
\newblock \href{https://doi.org/10.1103/PhysRevD.111.L081504}{doi:10.1103/PhysRevD.111.L081504}, \href{https://arxiv.org/abs/2407.10168}{arXiv:2407.10168}.

\bibitem{Zhang2025NoCauchy}
Cong Zhang, Jerzy Lewandowski, Yongge Ma, and Jinsong Yang.
\newblock Black holes and covariance in effective quantum gravity: a solution without {C}auchy horizons.
\newblock \emph{Phys. Rev. D}, 112:044054, 2025.
\newblock \href{https://doi.org/10.1103/PhysRevD.112.044054}{doi:10.1103/PhysRevD.112.044054},
\newblock \href{https://arxiv.org/abs/2412.02487}{arXiv:2412.02487}.

\bibitem{Belfaqih2025}
Idrus H. Belfaqih, M.~Bojowald, S.~Brahma, and E.~I. Duque.
\newblock Black holes in effective loop quantum gravity: covariant holonomy modifications.
\newblock \emph{Phys. Rev. D}, 112:046022, 2025.
\newblock \href{https://doi.org/10.1103/PhysRevD.112.046022}{doi:10.1103/PhysRevD.112.046022},
\newblock \href{https://arxiv.org/abs/2407.12087}{arXiv:2407.12087},

\bibitem{Zhang2025CovDyn}
Cong Zhang and Zhoujian Cao.
\newblock Covariant dynamics from static spherically symmetric geometries.
\newblock \emph{Phys. Rev. Lett.}, 135:261401, 2025.
\newblock \href{https://doi.org/10.1103/PhysRevLett.135.261401}{doi:10.1103/PhysRevLett.135.261401},
\newblock \href{https://arxiv.org/abs/2506.09540}{arXiv:2506.09540},

\bibitem{Yang2025Electrovacuum}
Jinsong Yang, Cong Zhang, and Yongge Ma.
\newblock Covariant effective spacetimes of spherically symmetric electrovacuum with a cosmological constant.
\newblock \emph{Phys. Rev. D}, 112:064049, 2025.
\newblock \href{https://doi.org/10.1103/PhysRevD.112.064049}{doi:10.1103/PhysRevD.112.064049},
\newblock \href{https://arxiv.org/abs/2503.15157}{arXiv:2503.15157},

\bibitem{Munch2023}
Johannes M{\"u}nch, Alejandro Perez, Simone Speziale, and Sami Viollet.
\newblock Generic features of a polymer quantum black hole.
\newblock \emph{Class. Quant. Grav.}, 40:135003, 2023.
\newblock \href{https://doi.org/10.1088/1361-6382/accccd}{doi:10.1088/1361-6382/accccd}, \href{https://arxiv.org/abs/2212.06708}{arXiv:2212.06708}.

\bibitem{Zhang2023LQBH}
Xiangdong Zhang.
\newblock Loop quantum black hole.
\newblock \emph{Universe}, 9:313, 2023.
\href{https://doi.org/10.3390/universe9070313}{\nolinkurl{doi:10.3390/universe9070313}}, \href{https://arxiv.org/abs/2308.10184}{\nolinkurl{arXiv:2308.10184}}.

\bibitem{Ashtekar2006Improved}
Abhay Ashtekar, Tomasz Pawlowski, and Parampreet Singh.
\newblock Quantum nature of the big bang: Improved dynamics.
\newblock \emph{Phys.\ Rev.\ D}, 74:084003, 2006.
\newblock \href{https://doi.org/10.1103/PhysRevD.74.084003}{doi:10.1103/PhysRevD.74.084003}, \href{https://arxiv.org/abs/gr-qc/0607039}{arXiv:gr-qc/0607039}.

\bibitem{Bojowald2008LRR}
Martin Bojowald.
\newblock Loop quantum cosmology.
\newblock \emph{Living Rev.\ Rel.}, 8:11, 2005.
\newblock \href{https://doi.org/10.12942/lrr-2005-11}{doi:10.12942/lrr-2005-11}, \href{https://arxiv.org/abs/gr-qc/0601085}{arXiv:gr-qc/0601085}.

\bibitem{HanLiu2020Improved}
Muxin Han and Hongguang Liu.
\newblock Improved $\bar{\mu}$-scheme effective dynamics of full loop quantum gravity.
\newblock \emph{Phys.\ Rev.\ D}, 102:064061, 2020.
\newblock \href{https://doi.org/10.1103/PhysRevD.102.064061}{doi:10.1103/PhysRevD.102.064061}, \href{https://arxiv.org/abs/1912.08668}{arXiv:1912.08668}.

\bibitem{Bojowald2007Lattice}
Martin Bojowald, Daniel Cartin, and Gaurav Khanna.
\newblock Lattice refining loop quantum cosmology, anisotropic models and stability.
\newblock \emph{Phys. Rev. D}, 76:064018, 2007.
\newblock \href{https://doi.org/10.1103/PhysRevD.76.064018}{doi:10.1103/PhysRevD.76.064018}, \href{https://arxiv.org/abs/0704.1137}{arXiv:0704.1137 [gr-qc]}.

\bibitem{Bojowald2006Inhom}
Martin Bojowald.
\newblock Loop quantum cosmology and inhomogeneities.
\newblock \emph{Gen. Rel. Grav.}, 38:1771--1795, 2006.
\newblock \href{https://doi.org/10.1007/s10714-006-0348-4}{doi:10.1007/s10714-006-0348-4}, \href{https://arxiv.org/abs/gr-qc/0609034}{arXiv:gr-qc/0609034}.

\bibitem{Falcke2000}
Heino Falcke, Fulvio Melia, and Eric Agol.
\newblock Viewing the shadow of the black hole at the galactic center.
\newblock \emph{Astrophys. J. Lett.}, 528:L13, 2000.
\newblock \href{https://doi.org/10.1086/312423}{doi:10.1086/312423}, \href{https://arxiv.org/abs/astro-ph/9912263}{arXiv:astro-ph/9912263}.

\bibitem{Johannsen2012}
Tim Johannsen and Dimitrios Psaltis.
\newblock Testing the no-hair theorem with observations in the electromagnetic spectrum: II. black-hole images.
\newblock \emph{Astrophys. J.}, 718:446--454, 2010.
\newblock \href{https://doi.org/10.1088/0004-637X/718/1/446}{doi:10.1088/0004-637X/718/1/446}, \href{https://arxiv.org/abs/1005.1931}{arXiv:1005.1931}.

\bibitem{Zhang2020}
Cong Zhang, Yongge Ma, Shupeng Song, and Xiangdong Zhang.
\newblock Loop quantum {S}chwarzschild interior and black hole remnant.
\newblock \emph{Phys.\ Rev.\ D}, 102:041502, 2020.
\newblock \href{https://doi.org/10.1103/PhysRevD.102.041502}{doi:10.1103/PhysRevD.102.041502}, \href{https://arxiv.org/abs/2006.08313}{arXiv:2006.08313}.

\bibitem{Zhang2022LQC}
Jerzy Lewandowski, Yongge Ma, Jinsong Yang, and Cong Zhang.
\newblock Quantum Oppenheimer-Snyder and Swiss Cheese models.
\newblock \emph{Phys. Rev. Lett.}, 130:101501, 2023.
\newblock \href{https://doi.org/10.1103/PhysRevLett.130.101501}{doi:10.1103/PhysRevLett.130.101501}, \href{https://arxiv.org/abs/2210.02253}{arXiv:2210.02253}.

\bibitem{Konoplya2025}
R. A. Konoplya and O. S. Stashko.
\newblock Probing the effective quantum gravity via quasinormal modes and shadows of black holes.
\newblock \emph{Phys. Rev. D}, 111:104055, 2025.
\newblock \href{https://doi.org/10.1103/PhysRevD.111.104055}{doi:10.1103/PhysRevD.111.104055}, \href{https://arxiv.org/abs/2408.02578}{arXiv:2408.02578}.

\bibitem{Chen2025}
Jiawei Chen and Jinsong Yang.
\newblock Shadows and optical appearance of quantum-corrected black holes illuminated by static thin accretions.
\newblock \emph{Eur. Phys. J. C}, 85:512, 2025.
\newblock \href{https://doi.org/10.1140/epjc/s10052-025-14230-w}{doi:10.1140/epjc/s10052-025-14230-w}.
\newblock \href{https://arxiv.org/abs/2503.06215}{arXiv:2503.06215}.

\bibitem{Lin2024}
Wentao Liu, Di Wu, and Jieci Wang.
\newblock Light rings and shadows of static black holes in effective quantum gravity.
\newblock \emph{Phys. Lett. B}, 858:139052, 2024.
\newblock \href{https://doi.org/10.1016/j.physletb.2024.139052}{doi:10.1016/j.physletb.2024.139052}, \href{https://arxiv.org/abs/2408.05569}{arXiv:2408.05569}.

\bibitem{Perlick2022}
Volker Perlick and Oleg Yu. Tsupko.
\newblock Calculating black hole shadows: Review of analytical studies.
\newblock \emph{Phys. Rep.}, 947:1--39, 2022.
\newblock \href{https://doi.org/10.1016/j.physrep.2021.10.004}{doi:10.1016/j.physrep.2021.10.004}, \href{https://arxiv.org/abs/2105.07101}{arXiv:2105.07101}.

\bibitem{Gralla2019}
Samuel E. Gralla, Daniel E. Holz, and Robert M. Wald.
\newblock Black hole shadows, photon rings, and lensing rings.
\newblock \emph{Phys. Rev. D}, 100:024018, 2019.
\newblock \href{https://doi.org/10.1103/PhysRevD.100.024018}{doi:10.1103/PhysRevD.100.024018}, \href{https://arxiv.org/abs/1906.00873}{arXiv:1906.00873}.

\bibitem{ChenYang2025b}
Jiawei Chen and Jinsong Yang.
\newblock Periodic orbits and gravitational waveforms in quantum-corrected black hole spacetimes.
\newblock \emph{Eur. Phys. J. C}, 85:726, 2025.
\newblock \href{https://doi.org/10.1140/epjc/s10052-025-14457-7}{doi:10.1140/epjc/s10052-025-14457-7}.
\newblock \href{https://arxiv.org/abs/2505.02660}{arXiv:2505.02660}.

\end{thebibliography}
\end{document}